\documentclass[a4paper,preprint,aps,pre]{revtex4-2}
\usepackage{graphicx}% Include figure files
\usepackage{dcolumn}% Align table columns on decimal point
\usepackage{bm}
\usepackage{color}
\usepackage{amssymb}
\usepackage{amsmath}
\usepackage{xfrac}
\usepackage{mathrsfs}
\usepackage{comment}
\usepackage{titlesec}
\renewcommand{\thesection}{\arabic{section}}
\renewcommand{\thesubsection}{\thesection.\arabic{subsection}}
\renewcommand{\thesubsubsection}{\thesubsection.\arabic{subsubsection}}
\def\br{{\bf r}}
\def\bu{{\bf \hat{u}}}
\def\bp{{\bf p}}

\begin{document}

\title{Hydrodynamics of passive and active rods in a temperature gradient}

\author{Anweshika Pattanayak}
\email{ph16036@iisermohali.ac.in}
\author{Abhishek Chaudhuri}
\email{abhishek@iisermohali.ac.in}
\affiliation{Department of Physical Sciences, Indian Institute of Science Education and Research Mohali, Sector 81, Knowledge City, S. A. S. Nagar, Manauli PO 140306, India}

\date{\today}

\begin{abstract}  
Temperature plays a very important role in various biological processes like the evolution of life, as it is anticipated that early life existed in a very hot environment that eventually cooled down with time. In vitro experiments, conducted with various active matter systems in anisotropic temperature environments, show several interesting outcomes. Motivated by these experiments and some simulation results, we study the collective behavior of self-propelled hard rods with excluded volume interactions, moving on a substrate in two dimensions and subjected to a small but finite external temperature gradient. Using a coarse-grained hydrodynamic approach, we explore the behavior of the system in the steady state and the stability both in the passive and active limits.
\end{abstract}

\maketitle
\section{Introduction}

The role of temperature in biological processes cannot be overstated. It forms the principal evolutionary driver of chemical reactions. Experiments have shown that early evolution happened faster due to the sensitivity between temperature and reaction rates and that catalytic efficiencies are significantly affected by environmental temperatures~\cite{stockbridge2010impact,nguyen2017evolutionary,xu2020recent,muchowska2020nonenzymatic,arcus2020temperature}. In the presence of a temperature gradient, polynucleotides such as RNA shows an escalation in polymerization rates~\cite{mast2013escalation,niether2016accumulation}. A similar experiment with tubulins which are monomeric units of microtubules, in the presence of a temperature gradient in a confined space,  showed polymerization of the tubulins, producing well-oriented microtubule assemblies with preferential polarity~\cite{kakugo2009formation}. Such accumulations of RNA or microtubules in spatially confined thermal gradient is attributed to the phenomenon of thermophoresis.  

Thermophoresis or the Soret effect deals with the drift or motion of particles of a fluid system along temperature gradients. It plays an important role in various applications like microfluidics~\cite{vigolo2010thermophoresis,yang2016thermoosmotic}, convection in ferrofluids~\cite{blums1998soret}, crude oil characterization~\cite{ghorayeb2003interpretation}, and the fabrication of synthetic microswimmers~\cite{jiang2010active,yang2014self}. Theoretical studies to understand linear and nonlinear effects in thermophoresis use either a hydrodynamic or thermodynamic viewpoint~\cite{burelbach2018unified,burelbach2019determining,fayolle2008thermophoresis,parola2004particle,ruckenstein1981can,wurger2013soret,bringuier2003colloid,duhr2006molecules,dhont2007thermodiffusion,rasuli2008soret}. However, several key questions remain about the nonlinear thermophoretic characteristics of particles and their dependence on particle shape and particle-solvent interactions and finally how these dependencies determine collective behavior~\cite{golestanian2012collective}.  

Most studies on thermophoresis look at spherical colloidal particles. However, filaments such as RNA or microtubules are elongated objects and naturally need to be treated differently. Microtubules can be modeled as passive rigid rods interacting via volume exclusion. Onsager showed that an ensemble of rigid rods undergoes a phase transition from a disordered state to a uniaxial orientational ordered state as a function of the density of the rods~\cite{onsager1949effects,doi1988theory}. Naturally, such systems show a rich variety of liquid crystalline phases. In a temperature gradient field, an anisotropic shape should be characterized by multiple thermal diffusivities leading to anisotropic thermophoresis~\cite{tan2017anisotropic}. Experimental and theoretical studies of anisotropic particles such as nematic liquid crystals in a temperature gradient report conflicting results about the orientation of the director relative to the gradient~\cite{stewart1936x,stewart1940orientation,currie1975orientation,yun1971thermal,patharkar1971interfacial}.  In a simulation study~\cite{sarman1994molecular,sarman2014director}, it was shown that the director of prolate ellipsoids aligns perpendicularly to the temperature gradient whereas the director of the oblate ellipsoids aligns parallel to the gradient. With a coupling between the temperature gradient and a possible orienting torque appears as quadratic in temperature gradient.  Simulations of colloidal rods in  a fluid, modeled with multiparticle collision dynamics, indicated that thermophoretic anisotropy due to the shape does not however show the rotation of the rods~\cite{tan2017anisotropic}. 

 The importance of thermophoresis is not limited to passive particles, but in the broader context of particles that self-propel~\cite{marchetti2013hydrodynamics,vicsek2012collective,cates2015motility,ramaswamy2010mechanics,zottl2016emergent}. For example, an experiment with silica beads half-coated with gold showed propulsion when irradiated with a defocused laser beam. The gold caps act as heat sources when they absorb light. Self-propelled particles serve as the prototype models for active matter, describing all living matter ranging from fish schools, bird flocks, and animal herds to cellular self-organization and bacterial aggregation~\cite{marchetti2013hydrodynamics,vicsek2012collective}. Rod-shaped bacteria~\cite{be2019statistical}, cytoskeletal filaments such as actin~\cite{schaller2010polar,huber2018emergence} and microtubules~\cite{sumino2012large}, shaken granular particles~\cite{aranson2006patterns,narayan2007long,kudrolli2008swarming,kumar2014flocking} and chemically driven rod-shaped Janus particles~\cite{paxton2006catalytically,walther2013janus} are example of active matter which are described as interacting self-propelled rods. These self-propelled rod-like particles show a rich variety of collective phenomena: formation of polar clusters~\cite{peruani2012collective,thutupalli2015directional}, long-ranged nematic order in fibroblasts~\cite{li2017mechanism}, counter-propagating density waves in multilayers of nematically ordered myxobacteria~\cite{igoshin2001pattern,borner2002rippling} and mesoscale turbulence in suspensions of swimming bacteria~\cite{wensink2012meso,sokolov2007concentration,sokolov2009reduction,dunkel2013fluid,benisty2015antibiotic}. Numerical studies of self-propelled rods show the formation of polar moving clusters and dynamic polar bands similar to the experimental systems~\cite{peruani2012collective,peruani2008mean,kraikivski2006enhanced,suzuki2015polar}. 

An approach to studying self-propelled rods in the absence of long-range hydrodynamic interactions (dry system) involves a continuum field theory approach derived from the microscopic dynamics of self-propelled rods~\cite{harvey2013continuum,peshkov2012nonlinear,baskaran2008enhanced,baskaran2008hydrodynamics,baskaran2010nonequilibrium,bertin2015comparison}. This approach provides a basis for identifying phases of self-propelled rods, the linear stability of patterns, associated phase transitions, and characterization of the full nonlinear dynamics at the macroscale. One such study of the dynamics of a collection of self-propelling hard rods~\cite{baskaran2008hydrodynamics} on a frictional substrate and interacting via hardcore collisions showed that while self-propulsion effectively tends to align colliding rods, the alignment is apolar i.e. particles align without distinguishing the head from the tail and the symmetry of the system remains nematic at large scales. 
%\textbf{How would such an ensemble of self-propelled rods behave in the presence of an external temperature gradient? Further, if the activity is switched off, how would the system of passive rods behave in the presence of this temperature gradient? }

To answer these questions, we study the collective behavior of self-propelled hard rods with excluded volume interactions, moving on a substrate in two dimensions and subjected to a small but finite external temperature gradient. We first establish that a rigid rod in a constant external temperature gradient will not experience torque. However, a torque in quadratic order is possible for a non-uniform temperature gradient. We show that an ensemble of passive rigid rods can align either perpendicular or parallel to a constant temperature gradient. The density field is inhomogenous and (increases)decreases as we move towards the hotter end for (thermophilic)thermophobic rigid rods. For an ensemble of active rods, the density field decays slower with increasing activity. For very high activity, the density field becomes homogeneous in the steady state, the density determining if the system is in the isotropic or the nematic state. 

%\end{tabular}
%\end{@twocolumnfalse}
\par

\section{Thermophoresis or Thermodiffusion for Rod-like Particles}

We first consider the effect of a temperature gradient on a single passive rod-shaped colloidal particle.  
%For a spherically symmetric colloidal particle, the velocity of the particle under the influence of a temperature gradient is given by ${\bf v}=-D^T{\bm{\nabla}}T$, where $D^T$ is the thermophoretic diffusion constant. 
Due to shape anisotropy, the components of translational velocity of the rod's center of mass parallel and perpendicular to its orientation are given as ${\bf v}_{||} = -D^T_{||}(\bm{\nabla}T\cdot\bu)\bu$ and ${\bf v}_{\perp} = -D^T_{\perp}\bm{\nabla}T\cdot({\bf \hat{I}}-\bu\bu)$ respectively, where $\bu$ is the unit vector of the long axis of the rod and $D^T_{||}, D^T_{\perp}$ are the parallel and perpendicular components of the diffusion tensor. The net translational velocity\cite{dhont2004thermodiffusion1,dhont2004thermodiffusion2,groot&mazur},
\begin{align}
	v_i&=-D^T_{ij} \partial_j T
\label{thermvel}
\end{align} 
with $D^T_{ij}=D^T_{||}u_iu_j+D^T_{\perp}(\delta_{ij}-u_iu_j)$. In the rest of the paper, we use $D^T_{\parallel}=D^T$ and $D^T_{\perp}=\alpha D^T_{\parallel}=\alpha D^T$ where $\alpha$ is a constant. 

Although the thermophoretic velocities are different for rod-shaped particles either aligned with the temperature gradient or perpendicular to it, a freely moving rod in a constant temperature gradient is not expected to experience a net torque as a thermophoretic force on one half of the rod is exactly the same as in the other half~\cite{tan2017anisotropic}. However, one may ask what happens if the temperature gradient is not a constant. Let us consider an infinitesimal length scale $dl$ at a distance $l$ from the rod's center of mass, $\br$. Drift velocity of the length segment,
\begin{align}
\begin{aligned}
{\bf v}(\br+l\bu) &= {\bf v}(\br)+l(\bu\cdot\bm{\nabla}){\bf v}+\frac{l^2}{2}u_i u_j\partial_i \partial_j {\bf v}+
\\ &...+\frac{l^n}{n!}u_{i1} u_{2i} ...u_{in}\partial_{i1}
\partial_{i2}...\partial_{in}\bf v+ ...
\end{aligned}
\end{align}
 The angular momentum of the rod is given as, 
\begin{align} 
\begin{aligned}
I\bm{\omega} &=\int_{-\frac{L}{2}}^
{\frac{L}{2}} l\bu\times {\bf v}(\br+l\bu) 
\\&= \bu\times {\bf v}(\br)\int_{-\frac{L}{2}}^
{\frac{L}{2}} l dl+ \bu\times (\bu\cdot\bm{\nabla}){\bf v}(\br)\int_{\frac{-L}{2}}^
{\frac{L}{2}} l^2 dl
\\&+\bu\times u_i u_j\partial_i \partial_j {\bf v}\int_{\frac{-L}{2}}^
{\frac{L}{2}}\frac{l^3}{2}dl+\bu\times u_{i1} u_{2i} ...u_{in}\partial_{i1}
\partial_{i2}...\partial_{in}{\bf v}\int_{\frac{-L}{2}}^
{\frac{L}{2}}\frac{l^{n+1}}{n!}dl+...
\end{aligned}
\end{align}
where $I$ is the moment of inertia of the rod with unit mass.
The angular velocity of the rod, $\bm{\omega}= \bu \times ((\bu\cdot\bm{\nabla})\bf{v}(\br) + \sum_{n=odd} a_n \partial_{i1}
\partial_{i2}...\partial_{in} {\bf {v}})$. Using Eq.~\ref{thermvel}, angular velocity can be written as,
\begin{align}
\omega^T_i=-D^T_{\perp} \epsilon_{ijk}u_ju_l\partial_l\partial_k T- D^T_{\perp}a_3\epsilon_{ijk}u_j u_l u_m u_n\partial_l\partial_m\partial_n\partial_k T+ ...
\end{align}
Therefore, a non-uniform temperature gradient can give rise to a torque in a rod-like colloid. 
\par
It is clear from the expression, that $\omega^T_i$ contains the terms with only even-order spatial derivatives of the temperature of the system. Hence for a unidirectional non-uniform temperature gradient, the nature of the angular momentum of a single particle is controlled by the temperature profile of the system. For a temperature profile that is even, the reversal of the direction of the temperature gradient does not affect the direction of thermophoretic torque and angular velocity. For an odd temperature profile, the reversal of the temperature gradient leads to the reversal of the direction of angular velocity.

\section{Smoluchowski equation}

We next consider the overdamped dynamics of an ensemble of rigid rods with volume exclusion, moving in two dimensions  on a substrate. Each rod of length $l$ and diameter $b$ ($l \gg b$), is characterized by the position of its center of mass ${\bf r}_i$ and its orientation $\bu_i = (\cos\theta_i, \sin\theta_i)$. A rod acts as though it were self-propelled by having a force ${\bf F}_i \propto v_0\bu_i$ operating on its center of mass and oriented along its long axis, where $v_0$ denotes the self-propulsion speed of the rod. Momentum exchange takes place with the substrate due to friction. In addition to the self-propulsion and volume exclusion, the rods are in the presence of an external temperature gradient $\bm{\nabla}T$. 

To describe the collective dynamics of the system of rigid rods we perform a noise average of the overdamped coupled Langevin equations of the position and orientation of the rods (see Supplementary \ref{Langevin_equation} for details). to give the Smoluchowski equation for the probability density function $\psi(\textbf{r},\bu,t)=\left\langle\Sigma_{j=1}^N\delta(\textbf{r}-\textbf{r}_j(t))\delta(\bu-\bu_j(t))\right\rangle$,
\begin{equation}
\partial_t \psi= -\bm{\nabla}\cdot{\textbf{J}}_t-\pmb{\mathscr{R}}\cdot\pmb{\mathscr{J}}_r
\label{smoluchowski}
\end{equation}
where the translational and rotational currents are given by $\textbf{J}_t$ and $\pmb{\mathscr{J}}_r$
respectively and $\pmb{\mathscr{R}} = \bu \times \frac{\partial}{\partial \bu}$ is the rotational operator~\cite{doi1988theory}.  The translational and rotational currents are the following : 
\begin{align}
\nonumber
{J}_{ti}&=-D_{ij}\nabla_j \psi-\frac{D_{ij}}{k_b T}\psi \nabla_j V_{ex}-D^T_{ij}\psi \nabla_j T+\psi v_0 u_i
\\
\mathcal{J}_{ri}&=-D_r\mathcal{R}_i \psi-\frac{D_r}{k_b T}\psi \mathcal{R}_i V_{ex}+\omega^T_i \psi
\label{currents}
\end{align}
where $D_{ij}=\frac{k_bT}{\gamma_{||}}u_iu_j+\frac{k_b T}{\gamma_{\perp}}(\delta_{ij}-u_iu_j) =D_{\parallel}u_iu_j+D_{\perp}(\delta_{ij}-u_iu_j) $ is the translational diffusion constant tensor and $D_r$ is the rotational diffusion constant. For a low density solution of long thin rods $D_{\perp}=D_{\parallel}/2$. For convenience, we choose $D_{\parallel} = D$ so that $D_{\perp} = D/2$ where $D=k_b T \ln(l/b)/2\pi\eta l$ and $D_r=6D/l^2$. As $D_{\parallel}>D_{\perp}$, the filaments are more likely to move along the orientation of the rod. For a small but finite temperature gradient, we have neglected the spatial variation of the diffusion coefficients( see Supplementary \ref{Langevin_equation}. 
The excluded volume interaction, $V_{ex}$ is obtained by calculating the probability of finding a rod within the interaction area of the other. In two dimensions for two thin rods characterized by $(\br_1, \bu_1)$ and $(\br_2, \bu_2)$, the excluded volume interaction is given as \cite{doi2013soft}
\begin{align}
\nonumber
V_{ex}(\br_1,\bu_1) & = 2 \nu k_B T \int d\br_2 \int d\bu_2 \psi(\br_2,\bu_2,t)|\bu_1 \times \bu_2 |
\\&\nonumber 
\times \int _{s_1,s_2} \delta(\br_1+\bu_1s_1-\br_2 - \bu_2s_2)
\\
& = 2 \nu k_B T \int d\bu_2\int _{s_1,s_2}\psi(\br_1+{\bm \xi}, \bu_2, t)|\bu_1 \times \bu_2 |
\label{excluded}
\end{align}
where  $-l/2\leq s_i\leq l/2$ parametrizes the position along the $i$-th filament, ${\bm{\xi}} = \br_2 - \br_1 = \bu_1s_1 - \bu_2s_2$ is the distance between the center of masses of the rods and $\nu$ is some arbitrary constant. The Dirac $\delta$ function is used to ensure that there is a point of contact within the interaction volume of the rods and  the excluded volume interaction potential works only if there is a contact.  $|\bu_1 \times \bu_2 |$ gives the excluded area of the two rods.

\section{Coarse-grained Smoluchowski Equation}

In this section, we will coarse-grain the Smoluchowski equation in order to obtain a set of equations that involve the macroscopic quantities on length scales large compared to the length of the rods. The macroscopic quantities are local density, $\rho(\br, t)$, the local polarization vector, $\bp(\br, t)$, and the second rank symmetric nematic alignment tensor, $S_{ij}(\br, t)$. These three quantities can be shown as the first three moments of the distribution $\psi(\br,\hat{u},t)$.
\begin{align}
\begin{aligned}
\rho(\br, t)&=\int_{\bu}\psi(\br, \bu, t)
\\
\rho{\bf p}(\br, t)&=\int_{\bu}\bu \psi(\br, \bu, t) = {\bf P}
\\
\rho S_{ij}(\br, t)&=\int_{\bu}\hat{Q}_{ij}\psi(\br, \bu, t) = Q_{ij}
\end{aligned}
\end{align}
where $\hat{Q}_{ij}(\bu) = u_iu_j - \frac{1}{2}\delta_{ij}$. Multiplying Eq.~\ref{smoluchowski} with $\bu$ and $\hat{Q}_{ij}$ and integrating with respect to $\bu$, we get the following dynamical equations for density, polarization vector, and alignment tensor respectively,
\begin{align}
\begin{aligned}
\partial_t \rho &= -\partial_i J_i
\\
\partial_t(\rho p_i)&=-\partial_jJ_{ij}-R_i
\\
\partial_t(\rho S_{ij})&=-\partial_k J_{ijk}-R_{ij}
\end{aligned}
\label{coarse_g_eqns}
\end{align}
where the translatioal currents are given by $J_i(\br,t) =\int_{\bu}J_{ti}(\br, \bu, t), J_{ij}(\br,t) = \int_{\bu} u_i J_{tj}(\br, \bu, t)$ and $J_{ijk}(\br, t) = \int_{\bu} \hat{Q}_{ij} J_{tk}(\br, \bu, t)$ respectively. The rotational currents, present in the equations for the polarization and alignment tensor are given as $R_i(\br, t) = \int_{\bu} \hat{u}_i {\pmb{\mathscr{R}}}\cdot{\pmb{\mathscr{J}}}_r(\br, \bu, t)$ and $R_{ij}(\br, t) = \int_{\bu} \hat{Q}_{ij} {\pmb{\mathscr{R}}}\cdot{\pmb{\mathscr{J}}}_r(\br, \bu, t)$ respectively. Note that the $J_{ti}$'s and $\pmb{\mathscr{J}}_{ri}$'s are given by Eq. ~\ref{currents}. 

The probability distribution $\psi(\br,\bu, t)$ can be expanded as a sum of irreducible tensors and written down in terms of moments of the distribution. We assume that at times longer than microscopic diffusion times, all the higher moments of the distribution become functionals of the first three moments. It then follows that in 2-dimensions. (see Supplementary \ref{moment_expansion} for details),
\begin{equation}
\psi(\br,\bu, t) = \frac{1}{2\pi}\rho(\br, t)\left[ 1 + 2{\bf p}(\br, t)\cdot\bu + 4S_{ij}(u_iu_j - \frac{1}{2}\delta_{ij})\right].
\end{equation}
Further, since we are interested in the long wavelength description of the system,$\psi(\br_1 + \bm{\xi},\bu_2, t)$ in Eq.~\ref{excluded}, can be expanded and truncated to second order in gradients. These two approximations facilitate the description of the system in terms of a closed set of macroscopic equations for the density, polarization, and nematic order parameters. In our work, we consider a small but constant temperature gradient along the $x-$axis(see Supplementary for the set of equations \ref{field_equations}). While the self-propulsion generates terms that couple the density and the nematic order parameter to the polarization even in the absence of excluded volume interactions, the temperature gradient couples the density and the nematic order parameter. We now analyze the steady state results.
%Therefore, both self-propulsion and temperature gradient creates a density flux in the direction of polarization.  

\section{Steady state results}
We choose a timescale $\tau \sim D_r^{-1}$ and a lengthscale $\sqrt{D/D_r}$. The equations can then be characterized by the two dimensionless quantities :  $\textrm{Pe} = v_0/\sqrt{DD_r}$ and the  Soret coefficient $S^T = D^T/D$.  We consider two separate cases: $(i)$ Passive rods in the absence and presence of a temperature gradient and $(ii)$ active rods in the absence and presence of a temperature gradient. 

%four different cases : $(i)$ Passive rods without temperature gradient : $Pe=0, \bm{\nabla} T = 0$; $(ii)$ Active rods without temperature gradient : $Pe \ne 0, \bm{\nabla} T = 0$; $(iii)$ Passive rods in temperature gradient : $Pe = 0, \vec{\nabla} T \ne 0$; and $(iv)$ Active rods in temperature gradient : $Pe \ne 0, \bm{\nabla} T \ne 0$.

\subsection{Ensemble of passive rods}
\subsubsection{In the absence of a temperature gradient}

In the absence of self-propulsion ($\textrm{Pe} = 0$) and a temperature gradient ($\bm{\nabla}T = 0$), the hydrodynamic equations are greatly simplified. The bulk states of the system are obtained from the solutions of homogeneous hydrodynamic equations. Dropping all the gradient terms and keeping the excluded volume interactions leads to the following equations
\begin{align}
    \begin{aligned}
       \partial_t \rho & = 0
       \\
        \partial_t p_i &=-\left(p_i- \nu l^2 \rho S_{ij}p_j \right)
        \\
        \partial_t S_{ij} &= -4\left(1-\rho\frac{\nu l^2}{4}\right)S_{ij}
       \label{passive}
    \end{aligned}
\end{align}
The homogeneous equations describe the equilibrium dynamics of overdamped hard rods with an isotropic liquid state at low density with $\rho = \rho_0$, ${\bf p} = 0$ and $S_{ij} = 0$. This state becomes unstable if $\rho > \rho_{in}$, where $\rho_{in} = \nu \frac{l^2}{4}$ is the critical density for the isotropic to nematic phase transition. For the nematic state, the alignment tensor becomes  $S_{ij}=S(n_in_j - \delta_{ij}/2)$. To get the steady state value of $S$, it is not sufficient to use Eq.~\ref{passive} and we need to incorporate higher order correction in the dynamical equation for $S_{ij}$. Further, the steady-state value of $p_i$ is zero in this case and it always remains stable, which is expected for passive apolar rods since orientations along $\bu$ and $-\bu$ are identical.

\begin{figure}[t]
\centering
  \includegraphics[scale=1.0]{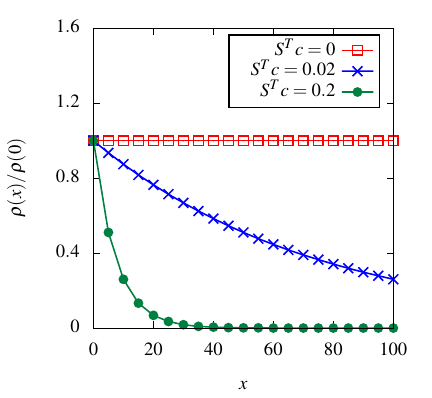}
  \caption{Spatial variation of density along $x$ for different temperature-gradient ($S^Tc$) values, plotted with the distance from the cool end of the system. Here, $\phi=\frac{\pi}{2},\alpha=0.3,S=0.8$ and $S^T$ is considered to be positive.}
  \label{stc}
\end{figure}

\subsubsection{In a temperature gradient}
As pointed out before, one of the key questions we wanted to address was the effect of an external temperature gradient on the collective behavior of anisotropic molecules. In this section, we look at the behavior of the rods in a temperature gradient ($\bm{\nabla}T \neq 0$) in the absence of an active self-propulsive force ($\textrm{Pe} = 0$). We have shown that an isolated rod does not experience torque in a constant temperature gradient. However, in a continuum setting, the density and the alignment tensor are coupled in the presence of a temperature gradient. In order to understand the collective behavior, we fix the direction of temperature gradient along the $x-$axis for simplicity and assume that $\bm{\nabla} T = c {\bf\hat{x}}$, where $c$ is a constant. Further, it is assumed that the spatial variations of the fields are along the direction of the temperature gradient. The time evolution equations for the density and the alignment tensor are simplified to
\begin{align}
\begin{aligned}
    \partial_t \rho &= \frac{1}{2}\partial_x^2Q_{xx}+\frac{3}{4}\partial_x^2 \rho +S^T(1-\alpha)\partial_x(Q_{xx}\partial_x T)
    \\&+\frac{S^T(1+\alpha)}{2}\partial_x(\rho\partial_x T)
    \\
     \partial_t Q_{xx}&= \frac{3}{4}\partial_x^2 Q_{xx}+\frac{1}{16}\partial_x^2\rho- 4\left(1-\rho\frac{\nu l^2}{4}\right)Q_{xx}
\\&+\frac{S^T(1+\alpha)}{2} \partial_x(Q_{xx}\partial_x T)+\frac{S^T(1-\alpha)}{8}\partial_x \rho \partial_x T
 \\
    \partial_t Q_{xy}&= \frac{3}{4}\partial_x^2 Q_{xy}- 4\left(1-\rho\frac{\nu l^2}{4}\right)Q_{xy}+\frac{S^T(1+\alpha)}{2} \partial_x(Q_{xy}\partial_xT) 
\end{aligned}
\end{align}
where $Q_{ij}=\rho S_{ij}$ with $S_{ij}=S\left(n_in_j-\frac{\delta_{ij}}{2}\right)$. Assuming that the alignment direction ${\bf\hat{n}}$ makes an angle $\phi$ with the direction of temperature gradient, we get $S_{xx}=\frac{S}{2}\cos(2\phi)$ and $S_{xy}=\frac{S}{2}\sin(2\phi)$. Now the fields are $\rho$, $S$, and $\phi$ and their dynamics are governed by the following equations.
\begin{align}
    %\begin{aligned}
\nonumber
     \partial _t \rho &=\frac{1}{4}\partial_x^2 \rho S \cos(2\phi)+S^Tc(1-\alpha)\frac{1}{2}\partial_x(\rho S \cos(2\phi)) \\&+\frac{3}{4}\partial_x^2 \rho + S^T c \frac{1+\alpha}{2}\partial_x(\rho)
\label{rhodyn}
     \\
\nonumber
      \rho S \partial_t \phi&=\frac{3}{4} (2\partial_x (\rho S)\partial_x\phi+\rho S\partial_x^2\phi)
      +S^T c\frac{1+\alpha}{2}S\left(  \rho \partial_x\phi\right)
      \\&
      -\sin(2\phi)\frac{1}{16}\left(\partial_x^2 \rho+2S^Tc(1-\alpha)\partial_x\rho \right)
\label{phidyn}
      \\  
\nonumber
      \partial_t (\rho S)& = \frac{\cos 2\phi}{8}\left(\partial_x^2 \rho+2 S^T c(1-\alpha)\partial_x \rho\right)+S^T c \frac{1+\alpha}{2}\partial_x(\rho S)\\&
-4\rho S\left(1 - \frac{\rho}{\rho_{in}}\right)+\frac{3}{4}\Big(\partial_x^2 (\rho S) -4 \rho S (\partial_x\phi)^2\Big) 
\label{Sdyn}
    %\end{aligned}
\end{align}

\begin{figure*}[t]
\centering
\includegraphics[scale=1.0]{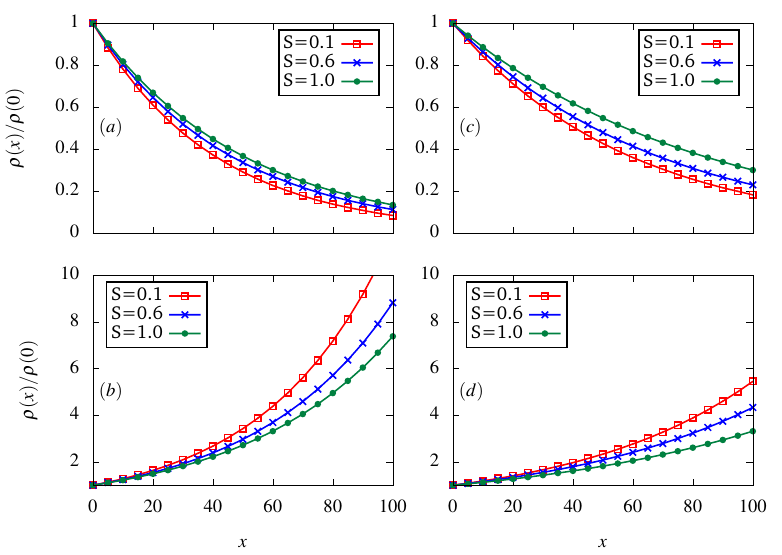}
\caption{Spatial variation of density for different $S$ values. 
\textbf{(a)}$\phi=0$, $\alpha=0.7$,$S^T c=0.02$ and $S^T>0$, \textbf{(b)}$\phi=0$, $\alpha=0.7$,$S^T c=0.02$ and $S^T<0$, \textbf{(c)}$\phi=\frac{\pi}{2}$, $\alpha=0.3$,$S^T c=0.02$ and $S^T>0$, \textbf{(d)}$\phi=\frac{\pi}{2}$, $\alpha=0.3$,$S^T c=0.02$ and $S^T<0$}
\label{changingS}
\end{figure*}

\paragraph{Steady state behaviour of density $\rho$}
In the stationary state,  $\partial_t \rho=0$ can be satisfied if (see Eq. \ref{rhodyn}), 
\begin{align}
\nonumber
    &\frac{1}{4}\partial_x (\rho S \cos(2\phi))+S^Tc(1-\alpha)\frac{1}{2}(\rho S \cos(2\phi)) +\frac{3}{4}\partial_x \rho\\& + S^T c \frac{1+\alpha}{2}(\rho)=0.
\label{rhosteady}
\end{align}
Neglecting spatial derivatives of $\phi$, i.e. assuming $\partial_x \phi \approx 0$, the steady state condition for $\phi$, $\partial_t \phi = 0$ gives $\sin(2\phi) = 0$. This implies that $\phi$ can take the following steady state values $\phi_{st} = 0, \pi/2$~\cite{khandkar2005orientational}. The implication of this result will be discussed in the next section. For these steady-state values,  Eq.~\ref{rhosteady} simplifies to give the density profile of the system as 
\begin{align}
    \rho_{st}(x)\vert_{\phi=0}&=\rho(0)\exp\left[\frac{-2S^Tcx\{1+\alpha+(1-\alpha)S\}}{3+S}\right]
    \label{steady_rho1}
      \\
    \rho_{st}(x)\vert_{\phi=\frac{\pi}{2}}&=\rho(0)\exp\left[\frac{-2S^Tcx\{1+\alpha-(1-\alpha)S\}}{3-S}\right]
    \label{steady_rho2}
\end{align}
where $x$ is the distance from the colder end of the system. In Fig.~\ref{stc}, we have plotted $\rho(x)$ for various values of the temperature gradient( specified by $S^Tc$), for the case $\phi = \pi/2$.  In the absence of a temperature gradient($S^Tc=0$), we have a homogenous steady state. As we increase the gradient, we observe the density falls off sharply towards the hotter end for thermophobic particles ($S^T>0$). 

If the particles were spherical in shape, then they would experience isotropic viscous force and thermophoretic force. Therefore, $D^T_{\parallel}=D^T_{\perp}=D^T$ and $D_{\parallel}=D_{\perp}=D$. In that case $\partial_t \rho=0$ would give, $D \bm{\nabla}\rho+D^T \rho\bm{\nabla} T=0$ and hence, $\bm{\nabla}\rho+S^T c \rho=0$. The density, as a function of temperature for such a system, is, $\rho(\br)=\rho(0)\exp\left[-S^T(T(\br)-T(0))\right]=\rho(0)\exp(-S^T c x)$ as obtained previously in \cite{duhrthermophoretic}.

%which gives density as a function of temperature, $\rho(\br)=\rho(0)\exp\left(-S^T(T(\br)-T(0))\right)=\rho(0)\exp(-S^T c x)$\cite{duhrthermophoretic}.

Note that we have considered the density $\rho(x)$ to be greater than the critical density required for the isotropic to nematic phase transition, i.e., we expect a uniform nematic state with $S>0$ and $\partial_x\phi\approx 0\approx \partial_x S $. From the plot of $\rho$ vs $x$ for different values of $S$ (see Fig.~\ref{changingS}), we can conclude that changing $S$ does not have much effect on the density for moderate values of $x$. Indeed, we can show the validity of this approximation by writing $S$ in terms of $\rho$ and $\partial_x \rho$ (see Supplementary for details). If we ignore the $S$ dependence in $\rho$, $\rho_{st}$ can be further simplified to have the same form for $\phi=0$, $\pi/2$ as 
 \begin{align}
 \nonumber
\rho_{st}(x)&\approx\rho(0)\exp\left(\frac{-2S^T c x(1+\alpha)}{3}\right)
\end{align}
In the limit of very small temperature gradient, 
\begin{align}
\rho_{st}(x)& \approx \rho(0)\left(1-\frac{2S^T c x(1+\alpha)}{3}\right)
\label{stableSpassive}
\end{align}
For $\alpha = 0.5$, this approximation is exact( see Eq \ref{steady_rho1},\ref{steady_rho2}). Then, for low-temperature gradients, the steady state density profile is linear and in the limit of vanishing temperature gradient, it gives a homogeneous distribution, as expected.

%\begin{align}
%\rho_{st}(x) = \rho_0\exp\left(\frac{-2S^T c x(1+\alpha)}{3}\right)
%\end{align}
\paragraph{Steady state behaviour of $\phi$ and $S$}
If we substitute this steady state density in the time evolution equation for $\phi$ (Eq.~\ref{phidyn}) and assume that spatial gradients in $\phi$ can be neglected then we get the following simplified dynamical equation for $\phi$, 
\begin{align}
\begin{aligned}
     \partial_t \phi&= \sin(2\phi)\frac{1}{6  S}{S^T}^2c^2(1-\alpha)(1+\alpha).
\end{aligned}
\label{phi1}
\end{align}
This tells us that the steady state values of $\phi = 0, \pi/2$ can be obtained provided $\alpha \neq 1$. We first note that these two steady state values of $phi$ imply that the rods can align parallel($\phi=0$) or perpendicular($\phi=\pi/2$) to the temperature gradient. To check for stability, we assume a small fluctuation about steady state value, i.e., $\delta \phi =\phi -\phi_{st}$. The linearized equation for the fluctuation in $\phi$ gives 
\begin{align}
\nonumber
    \partial_t \delta\phi&=  \frac{3}{4\rho_{st} S_{st}} (2{\partial_x(S_{st} \rho_{st}})\partial_x\delta\phi+\rho_{st} S_{st} \partial_x^2\delta\phi) + S^T\frac{1+\alpha}{2}\partial_x T\partial_x\delta\phi
     \\&
    -\frac{\cos(2\phi_{st})}{8 S_{st} \rho_{st}}\left[2S^T(1-\alpha)\partial_x\rho_{st}\partial_x T)\right]\delta\phi
\end{align}
Expanding $\delta\phi$ in Fourier components $\delta\phi=\sum_{k=0}^{\infty}\phi_k e^{-i\Vec{k}.\br}$,  gives
\begin{align}
%\begin{aligned}
\nonumber
      \partial_t \phi_k&=-  \left(\frac{3}{2} (i2k_x\partial_x \rho_{st} S_{st})+
      \frac{3}{4}k^2
      +i k_x S^T c\frac{1+\alpha}{2}\left( S \rho_{st}  \right)\right)\phi_k
      \\&
     - \frac{\cos(2\phi_{st})}{8 S_{st} \rho_{st}}\left(2S^Tc\left((1-\alpha)\partial_x\rho_{st}\right)\right)\phi_k=Z_k\phi_k
 %    \end{aligned}
  \end{align}
  The real part of $Z_k$ determines the stability of the steady state. 
  \begin{align}
    Re(z_k)\vert_{\phi=0}&=-\frac{3}{4}k^2+\frac{1}{6  S_{st}}{S^T}^2c^2(1-\alpha)(1+\alpha)=f_1
    \\
     Re(z_k)\vert_{\phi=\frac{\pi}{2}}&=-\frac{3}{4}k^2-\frac{1}{6  S_{st}}{S^T}^2c^2(1-\alpha)(1+\alpha)=f_2
\end{align}
Both $f_1$ and $f_2$ can be negative for high values of $k$. In the hydrodynamic limit, with $k\rightarrow 0$,  stability demands that $(i)$  $f_1 < 0$ if $\alpha > 1$ and $(ii)$ $f_2 < 0$ if $\alpha < 1$ where $\alpha = D^T_{\perp}/D^T_{\parallel}$.   

This additional restriction implies that alignment of the rods parallel ($\phi=0$) to the direction of the temperature gradient is stable if $D^T_\perp>D^T_\parallel$  and alignment of the rods perpendicular ($\phi=\pi/2$) to the direction of the temperature gradient is stable if $D^T_\perp<D^T_\parallel$. This is consistent with earlier molecular dynamics study with liquid crystals in the presence of an external temperature gradient where it was shown that $\phi=0$ is stable for oblate liquid crystals and $\phi=\pi/2$ is stable for prolate liquid crystals \cite{sarman1994molecular}.  
 
Using Eq. \ref{rhodyn}, the dynamical equation for $S$ \ref{Sdyn} can be rewritten as,
\begin{align}
    \begin{aligned}
        \partial_t S & =\frac{3}{4}\partial_x^2S +\frac{3}{2}\frac{\partial_x \rho}{\rho}\partial_x S -4\left(1-\frac{\rho_{st}}{\rho_{in}}\right) S\\& 
        +\frac{\cos 2\phi}{8}(1-2 S^2)\left(2 S^T(1-\alpha)\partial_x T\frac{\partial_x \rho}{\rho}\right)
        +S^T c \frac{1+\alpha}{2}\partial_x S
        \label{passive_Sdynamical}
    \end{aligned}
\end{align}
To solve the above equation for $\partial_t S=0$, we neglect terms such as $\partial_x^2S$ and $S^2$ and consider, $\rho_{st}(0)=\rho_{in}(1+\epsilon)$ for $S^T>0W$ i.e. the steady state density at $x=0$ is slightly above (or below, $\epsilon<0$) the critical density necessary for isotropic-nematic transition. Then we get,
\begin{align}
    \begin{aligned}
       & S^Tc(1+\alpha)\partial_x S_{st}+S_{st}\left(\frac{16}{3}S^Tc(1+\alpha)(1+\epsilon)x-8\epsilon\right)
       \\ & +\cos(2\phi)\frac{(S^Tc)^2(1-\alpha^2)}{3}=0
\label{Ssteady}
    \end{aligned}
\end{align}
We consider $\alpha>1$ for $\phi=0$ and $\alpha<1$ for $\phi=\pi/2$ as these are the stable states.  We can now solve for the above differential equation with the boundary condition that $S_{st}(x=0)=S_0$.  The solution for $S_{st}$ gives an error function (see Supplementary for details) which we truncate up to the linear order.  Further, for small values of $c$ (small temperature gradients) and $\epsilon$ (steady state densities are close to the isotropic-nematic critical point but firmly in the nematic regime), we obtain the following solution for the steady state orientational order parameter, 
\begin{align}
    \begin{aligned}
     S_{st}(x)&=\left[S_0\exp \left(\frac{6 \epsilon ^2}{(\alpha +1)^2 (S^Tc)^2 (\epsilon +1)}\right)+\frac{(1-\alpha ) S^T c x}{3}\right]
     \\& \exp \left[-\frac{6 \epsilon ^2}{(\alpha +1)^2 (S^Tc)^2 (\epsilon +1)}+\frac{8 \epsilon  x}{(\alpha  +1)S^Tc}-\frac{8 (\epsilon +1)}{3}  x^2\right]
    \end{aligned}
\end{align}

\begin{figure}[t]
\centering
\includegraphics[scale=0.8]{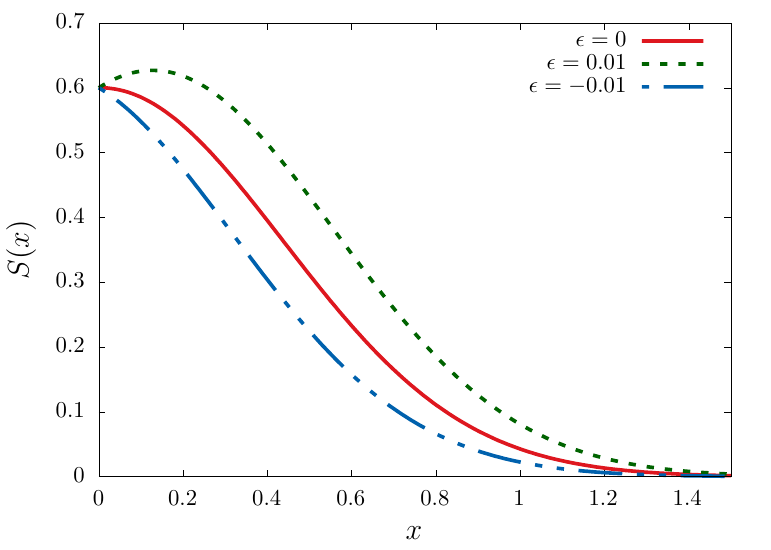}
\caption{Plot of steady-state $S(x)$ vs $x$ for $S^Tc=0.08$ and different $\epsilon$ values. We have fixed $\alpha=0.5$,$\phi=\pi/2$ and $S_0=0.6$. $S^T$ is considered to be positive.}
\label{variableS}
\end{figure}

%\begin{figure*}
%     \scalebox{.8}{\input{variableSnew}}
%\end{figure*}
$S_{st}(x)$ must be bounded in the region $[0,1]$ since $0 < S_0 < 1$ and we expect $S_{st}(x)$ to decrease monotonically with $x$. Further, we need to consider small temperature gradients (low $S^Tc$) where Eq.~\ref{stableSpassive} remains valid. In Fig.~\ref{variableS}, we plot $S_{st}(x)$ for fixed $\alpha, S_0$, and $S^Tc$ for two values of $\epsilon$. We note that at low temperature gradients, $S_{st}$ increases at small $x$ for some values of $\epsilon$. At higher values of $x$, $S_{st}$ decreases monotonically. The approximated solution suggests an enhanced nematic ordering as we move from the colder end for small $x$. We cannot comment if this hump will persist in a full numerical solution of the dynamical equations.

It is interesting to note that for some range of $\epsilon$, the effect of excluded volume dominates. In that limit, we can neglect the $\partial_x S$ term and the constant term in Eq. \ref{Ssteady}, for a small temperature gradient, leading to a steady state value of $S_{st}=0$ which can be unstable if the coefficient of $S$ in the equation of $\partial_t S$(Eq \ref{passive_Sdynamical}) is positive. This implies a nematic state. To get a non-zero value of $S_{st}$ we must incorporate higher order terms in $S$. For $S^T>0$(thermophobic particles), this approximation works best in the range of moderate $x-$ values.

\subsection{Ensemble of Active rods} 
\subsubsection{In the absence of a temperature gradient}
The collective behavior of self-propelled hard rods without a temperature gradient using coarse-grained hydrodynamic equations in the slow variables has been discussed earlier~\cite{baskaran2008enhanced}.  Here we will report the results without going into details. Although the self-propulsion breaks the nematic symmetry in the microscopic equations, it does not generate a macroscopic polarized state. The only bulk states of the self-propelled system are the ones in equilibrium: isotropic and nematic. The nature of fluctuations in the steady state shows the effect of self-propulsion giving rise to propagating waves in a range of wave vectors. Further, the nematic state is unstable above a critical $\textrm{Pe}$.

\subsubsection{In the presence of a temperature gradient}  

The density field for self-propelled rods in a constant temperature gradient only in the $x-$direction evolves as,
  \begin{align}
  \begin{aligned}
       \partial _t \rho &=\Big(\frac{1}{2}\partial_x^2 Q_{xx}+S^T(1-\alpha)\partial_x(Q_{xx}\partial_x T)\Big)+\frac{3}{4}\partial_x^2 \rho\\& + S^T \frac{1+\alpha}{2}\partial_x(\rho\partial_x T)-\textrm{Pe} \partial_x P_x
  \end{aligned}
  \end{align}
  As described earlier, $\textrm{Pe}=\sqrt{v_0^2/D D_r}$ is the dimensionless Peclet number. In the stationary state with $\partial_t \rho = 0$,
  \begin{align}
  \begin{aligned}
    & \frac{1}{2}\partial_x Q_{xx}+S^T(1-\alpha)Q_{xx}\partial_x T+\frac{3}{4}\partial_x \rho + S^T \frac{1+\alpha}{2}\rho\partial_x T-\textrm{Pe}P_x=0
      \label{densityss}
      \end{aligned}
  \end{align}
As before, using the director ${\bf{\hat n}}$ makes an angle $\phi$ with the direction of the temperature gradient, we get,
  \begin{align}
  \begin{aligned}
   &\left(\frac{3}{2}+\cos(2\phi)\frac{S}{2}\right)\partial_x \rho+S^T(1-\alpha)\cos(2 \phi)\rho S \partial_x T\\&+S^T(1+\alpha)\rho\partial_x T-(1+2 \textrm{Pe}^2)\sin(2\phi)\rho\partial_x \phi
   -\textrm{Pe} \rho p_x=0
   \label{rhoSactive}
   \end{aligned}
   \end{align}
where the last term can be replaced from the following equation of polarization vector in steady state,
   \begin{align}
   & \rho p_x = -\textrm{Pe}\left(\frac{1}{2}\partial_x \rho+\frac{1}{2}\partial_x \rho S \cos(2\phi)\right)
  \end{align}

%\end{multicols}

The $\phi$ field evolves as,
  \begin{align}
      \begin{aligned}
          \rho S \partial_t \phi&=\frac{3+\textrm{Pe}^2}{4} (2\partial_x \rho\partial_x\phi+2 \rho\partial_x^2\phi)S
      +S^T\frac{1+\alpha}{2}\left( S \rho \partial_x\phi\partial_x T \right) \\&
      -\frac{\sin(2\phi)}{16}\left[ \partial_x^2 \rho(1+2 \textrm{Pe}^2)+2S^T((1-\alpha)\partial_x\rho\partial_x T\right]
      \label{phi}
      \end{aligned}
  \end{align}
 % \begin{multicols}{2}
    For a nematic steady state, $\partial_t \phi =0$. Ignoring spatial variations of $\phi$ ( $\partial_x\phi \approx 0$ and $\partial^2_x \phi \approx 0$), it is easy to see that the steady state conditions for $\phi$ under such circumstances are $\phi=0$ and $\phi=\frac{\pi}{2}$ as in the passive case. Putting these $\phi$'s in Eq.~\ref{rhoSactive} we can derive the following expressions for the density fields :
  \begin{align}
      \rho_{st}(x)\vert_{\phi=0}&=\rho(0)\exp\left(\frac{-2S^Tcx\{1+\alpha+(1-\alpha)S\}}{3+S+2\textrm{Pe}^2(1+S)}\right)
      \\
      \rho_{st}(x)\vert_{\phi=\frac{\pi}{2}}&=\rho(0)\exp\left(\frac{-2S^Tcx\{1+\alpha-(1-\alpha)S\}c x}{3-S+2\textrm{Pe}^2(1-S)}\right).
  \end{align}
  Since we have ignored spatial variations of $S$, in the limit of small $S$, these density fields for both the $\phi$ values reduce to the same form : 
  \begin{flalign}
      \rho_{st}(x)&\approx\rho(0)\exp\left(\frac{-2S^T c x(1+\alpha)}{3+2 \textrm{Pe}^2}\right)
      %\label{active_rho}
  \end{flalign}
  If the temperature gradient is small i.e. small $S^Tc$, $\rho_{st}$ can be further approximated to 
  \begin{flalign}
      \rho_{st}(x)&\approx \rho(0)\left(1-\frac{2S^T c x(1+\alpha)}{3+2 \textrm{Pe}^2}\right)
      \label{active_rho}
  \end{flalign}
  For the passive case($Pe=0$), we retrieve the earlier result (Eq. \ref{steady_rho1},\ref{steady_rho2}). At low $\textrm{Pe}$, $\rho_{st}$ falls sharply with $x$from the density at $x=0$. At high $\textrm{Pe}$, the fall is minimal as the rods assume the homogeneous density($\rho_0$) even in the presence of the temperature gradient. 

\begin{figure}[t]
\centering
\includegraphics[scale=1.0]{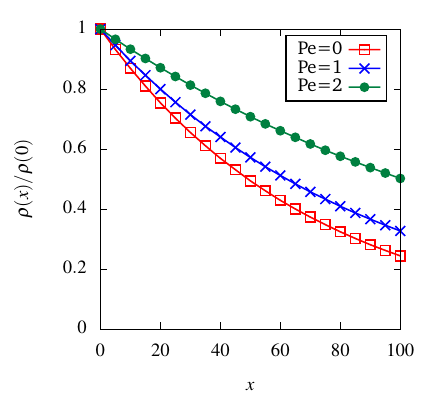}
\caption{Spatial variation of density for varying activity for thermophoretic particles($S^T>0$). }
\label{changingPe}
\end{figure}

%\end{minipage}
  \subsubsection{Stability of steady states} 
  Following the same procedure as for the passive rods, neglecting spatial variations of $\phi$ and using the expression of $\rho(x)$ as given by equation \ref{active_rho}, the simplified dynamical equation for $\phi$ gives,
  \begin{align}
      \begin{aligned}
          \partial_t \phi&=
      \frac{\sin(2\phi)}{12}\left((S^Tc)^2(1-\alpha^2)\right)
      \end{aligned}
  \end{align}
  Therefore for a nematic steady state, $\phi$ can have two stable solutions, $0$ and $\pi/2$. 
  To check the stability of $\phi$, we consider fluctuations $\delta \phi =\phi -\phi_{st}$ from its steady state. The dynamic equation of $\phi$ is written up to the linear order of $\delta\phi$
  % \end{multicols}
  \begin{align}
      \begin{aligned}
%          \rho_{st} S \partial_t \delta\phi&=\frac{3+Pe^2}{4} (2\partial_x \rho_{st}\partial_x\phi_{st}+ 2\rho_{st}\partial_x^2\delta\phi)S
%      +S^T c\frac{1+\alpha}{2}\left( S \rho_{st} \partial_x\delta\phi \right) \\ &
%      -\frac{\sin(2\phi_{st})}{16}\partial_\phi\left(2S^T c\left((1-\alpha)\partial_x\rho\right)\right)\Big\vert_{st}\delta\phi
%       \\ &-\frac{\cos(2\phi_{st})}{8}\left(2S^T\left((1-\alpha)\partial_x\rho\partial_x T\right)\right)\Big\vert_{st}\phi
%      \\ \Rightarrow 
\partial_t \delta\phi&=  \frac{(3+\textrm{Pe}^2)}{4\rho_{st}} (2{\partial_x \rho_{st}}\partial_x\delta\phi+ \partial_x^2\delta\phi) + S^T c\frac{1+\alpha}{2}\partial_x\delta\phi \\
      &-\frac{\cos(2\phi_{st})}{8 S \rho_{st}}\left[2S^T c\left((1-\alpha)\partial_x\rho_{st}\right)\right]\delta\phi
      \label{phif}
      \end{aligned}
  \end{align}
  %\begin{multicols}{2}
   Expanding the fluctuation as a Fourier series, $\delta\phi=\sum_{k=0}^{\infty}\phi_k e^{-i\Vec{k}.\br}$, the linearized equation for $k-$th Fourier mode is obtained as follows :
  \begin{align}
  \begin{aligned}
      \partial_t \phi_k&=  \left(-\frac{3+Pe^2}{4} (i2k_x\frac{\partial_x \rho_{st}}{\rho_{st}}+k^2)
      -i k_x S^Tc\frac{1+\alpha}{2}\right.
      \\
     & \left.+\cos(2\phi_{st})\frac{(S^T c)^2(1-\alpha^2)}{6 S }\right)\
  \end{aligned}
  \end{align}
  Time evolution of $\phi_k$ is governed by a frequency $z(k)$ i.e. $\partial_t \phi_k=z(k)\phi_k$. Real part $z(k)$ determines the stability of the steady states.
Using expressions of $\rho_{st}$'s 
\begin{align}
    Re(z_k)\vert_{\phi=0}&=-k^2\frac{3+Pe^2}{4}+\frac{\left((S^T c)^2(1-\alpha^2)\right)}{(3+2 Pe^2) S }=f_1
    \\
     Re(z_k)\vert_{\phi=\frac{\pi}{2}}&=-k^2\frac{3+Pe^2}{4}-\frac{\left((S^T c)^2(1-\alpha^2)\right)}{(3+2 Pe^2) S }=f_2
\end{align}

For small or moderate values of $\textrm{Pe}$, both $f_1$ and $f_2$ can be negative as in the case of passive rods and therefore give stable solutions, for large values of $k$, where the first term will dominate. In the hydrodynamic limit ($k\rightarrow 0$) however,  we can have $(i) f_1 < 0$ provided $\alpha > 1$ and $(ii) f_2 < 0$ provided $\alpha < 1$. Therefore, just like the passive rods, self-propelled rods tend to align perpendicular to the direction of the temperature gradient if $D^T_\perp<D^T_\parallel$ and parallel ($\phi=0$) to the direction of the temperature gradient if $D^T_\perp>D^T_\parallel$. For large $\textrm{Pe}$, $Re(z_k)\vert_{\phi=0}=Re(z_k)\vert_{\phi=\frac{\pi}{2}}=-\frac{\textrm{Pe}^2 k^2}{4}$ and either state is stable.
%Interestingly, however, an increase in $\textrm{Pe}$ leads to greater stability of the $\phi = 0$ and $\pi/2$ phases. 
    %So, alignment perpendicular to the direction of the temperature gradient is stable if $\alpha<\frac{1}{2(1+Pe^2)}$ which is very low if $Pe$ is high. Alignment parallel to temperature gradient shows opposite behavior. 
    \par

  At steady state, when $\partial_t \rho=0$ and ignoring higher order spatial variations in $\rho$ ($\partial_x^2 \rho\approx 0$),  the dynamical equation satisfied by $S$ is,
   \begin{align}
  \begin{aligned}
      \partial_t S& = (1-2 S^2)\frac{\cos 2\phi}{8}\left[2 S^T(1-\alpha)c\frac{\partial_x \rho}{\rho}\right]-4\left(1-\frac{\rho}{\rho_{in}}\right) S \\
      &+\frac{S^T c(1+\alpha)}{2}\partial_x S +\left(\frac{3+\textrm{Pe}^2}{2}\right)\left[\partial_x S\frac{\partial_x\rho}{\rho}\right]  
  \end{aligned}
  \end{align}
   Here we have neglected the $S^2$ term. Again, as before if we consider $\rho(0)=\rho_{in}(1+\epsilon)$ then $\partial_t S=0$ gives, 
  \begin{align}
    \begin{aligned}
       & \frac{S^Tc(1+\alpha)}{2}\left(1-\frac{2(3+\textrm{Pe}^2)}{3+2\textrm{Pe}^2}\right)\partial_x S_{st}
       \\&-4 S_{st}\left(\frac{2S^Tc(1+\alpha)(1+\epsilon)x}{3+2\textrm{Pe}^2}-\epsilon\right)-\cos(2\phi)\frac{(S^Tc)^2(1-\alpha^2)}{2(3+2\textrm{Pe}^2)}=0
       \label{active_steady_S}
    \end{aligned}
\end{align}
We consider three different regimes of $Pe$ : 
\begin{enumerate}
    \item $\mathbf{Pe<<1}$: In this limit, $\textrm{Pe}^2$ is negligible and the equation reduces to the same as that for the passive rods (see Eq.~\ref{Ssteady}).
    \item $\mathbf{Pe\approx1}$: In this limit, equation \ref{active_steady_S} takes the form  
    \begin{align*}
    \begin{aligned}
       & \frac{3 S^Tc(1+\alpha)}{3+2\textrm{Pe}^2}\partial_x S_{st}+8 S_{st}\left(\frac{2S^Tc(1+\alpha)(1+\epsilon)x}{3+2\textrm{Pe}^2}-\epsilon\right)
       \\ & +\cos(2\phi)\frac{(S^Tc)^2(1-\alpha^2)}{(3+2\textrm{Pe}^2)}=0       
    \end{aligned}
\end{align*}
In this case, the co-efficient of $\partial_xS$ and the constant term are lower than in the passive rods case($\textrm{Pe}=0$). Hence we expect a significant spatial variation of $S$ for lesser values of $\epsilon$'s than that for the passive rods.
\item $\mathbf{Pe>>1}$: In this limit $\frac{2(3+\textrm{Pe}^2)}{3+2\textrm{Pe}^2} \approx 1$ and the co-efficient of $\partial_x S$ vanishes. The coefficient of the $\cos(2\phi)$ term is extremely small and can be neglected. This leads to the following equation,
%\begin{align*}
%    \frac{S^Tc(1+\alpha)}{2}\left(1-\frac{2(3+\textrm{Pe}^2)}{3+2\textrm{Pe}^2}\right)\approx 0
%\end{align*}
%Similarly, we can neglect the constant  $+cos(2\phi)\frac{(S^Tc)^2(1-\alpha^2)}{(3+2Pe^2)}$ for small $c$ values. Hence, in high $Pe$ limit, equation \ref{active_steady_S} becomes 
\begin{align}
   S_{st}\left(\frac{2S^Tc(1+\alpha)(1+\epsilon)x}{3+2\textrm{Pe}^2}-\epsilon\right)=0
\end{align}
which gives only one steady state value, $S_{st}=0$. If we look at the steady state equation for $S$, it takes the form $\partial_t S=-aS$ which implies that $S_{st} = 0$ becomes unstable for $a < 0$ leading to an isotropic-nematic transition. This again implies a shift towards homogeneity.
%Dynamical equation Of $S$ becomes $\partial_t S=-a S$, implying that $S_{st}=0$ becomes unstable for $a<0$ leading to an isotropic-nematic transition.
\end{enumerate}
The self-propulsion speed induces a collective polarization. The simplest expressions of the components of the polarization vector in the steady state are,
 \begin{align}
      \begin{aligned}
      p_x &= \frac{\textrm{Pe}}{2}(1+S\cos 2\phi ) \frac{\partial_x \rho}{\rho}
      \\
       & \approx -\frac{\textrm{Pe}}{2}\left(1\pm S \right)\Big(\frac{2S^Tc(1+\alpha)}{(3+2 \textrm{Pe}^2)}\Big)
      \\
      p_y &= \frac{\textrm{Pe}}{2}(S \sin 2\phi) \frac{\partial_x \rho}{\rho}=0
      %\textbf{p} &= -\frac{Pe}{2}\left(1\pm S \right)\Big(\frac{2S^Tc(1+\alpha)}{(3+2 \textrm{Pe}^2)}\Big)\left(1+x\frac{2S^Tc(1+\alpha)}{(3+2 \textrm{Pe}^2)}\right)\hat{\textbf{x}}
      \end{aligned}
  \end{align}
The polarization vector is therefore directed opposite to the direction of the temperature gradient. This has been observed in \cite{kakugo2009formation}, where polarities of the microtubules were mostly directed towards the colder end. For large $\textrm{Pe}$, $\textbf{p}=0$, which again implies the absence of inhomogeneity.
%and $\vert p_x\vert $ increases with increasing $x$ ($\vert\partial_x p_x\vert$ is small ) indicating that the system is more polarized in the warm end. 
\begin{comment}    
   \begin{figure*}
       \scalebox{1.4}{\input{steady_S0}}
   \end{figure*}
\end{comment}
\section{Discussions}
In this paper, we have studied the behavior of rod-like particles in the presence of a small but finite external temperature gradient. We assume that each particle feels a thermophoretic force acting on its center of mass because of the temperature gradient, giving it a translational drift velocity. Although the component of the translational drift velocity along the orientation of the rod is different from the component perpendicular to the orientation of the rod, a uniform temperature gradient cannot induce any torque on the particle. A non-uniform temperature gradient can induce a torque, as observed in earlier simulations. 

We studied the collective behavior of such rod-like particles interacting with each other via excluded volume interaction, by coarse-graining the Smoluchowski equation in the presence of a uniform temperature gradient. Unlike the single-particle case, the ensemble of rod-like particles can orient collectively due to the coupling of the density and nematic order parameter field to the temperature gradient. The stable steady state direction of the alignment is either parallel or perpendicular to the direction of the temperature gradient. In the steady state, we have shown that if we go from the cooler to the hotter end of the system, the density of the particles either decay (thermophobic) or increase (thermophilic), with the rate of decay depending on the temperature gradient and activity. A linear stability analysis also shows that if the thermophoretic drift in the direction parallel to the orientation of the particle is more (less) than the thermophoretic drift in the direction perpendicular to it, the particles prefer to orient perpendicular (parallel) to the temperature gradient. 

Next, we analyzed the steady state behavior of the nematic order parameter for thermophobic particles only. If the density of the particles is higher than the critical density necessary for the isotropic-nematic phase transition in the absence of a temperature gradient, we expect a uniform non-zero nematic order all over the space. Interestingly, even if the density is lower than the critical density, we observe a finite nematic order when there is a temperature gradient. The nematic order parameter decays to zero if we go toward the hotter end. So, however small, the temperature gradient can have an effect on the collective orientation of the rod-like particles. 

In the case of the active rod-like particles, the density of the particles becomes more and more homogeneous if we increase the activity of the particles implying that self-propulsion drift and diffusion dominate over thermophoresis. The analysis of the nematic order parameter reflects the same result. Activity leads to an average polarization directed opposite to the temperature gradient.

\section*{Author Contributions}
A.P. carried out this work, and A.C. planned this work. 
%We strongly encourage authors to include author contributions and recommend using \href{https://casrai.org/credit/}{CRediT} for standardized contribution descriptions. Please refer to our general \href{https://www.rsc.org/journals-books-databases/journal-authors-reviewers/author-responsibilities/}{author guidelines} for more information about authorship.

\section*{Conflicts of interest}
There are no conflicts to declare.

\section*{Acknowledgements}
We acknowledge Arnab Saha for a careful reading of the manuscript.

%%%END OF MAIN TEXT%%%

%The \balance command can be used to balance the columns on the final page if desired. It should be placed anywhere within the first column of the last page.

%\balance

%If notes are included in your references you can change the title from 'References' to 'Notes and references' using the following command:
%\renewcommand\refname{Notes and references}

%%%REFERENCES%%%
\bibliography{reference} %You need to replace "rsc" on this line with the name of your .bib file
\bibliographystyle{apsrev4-2} %the RSC's .bst file
%\end{multicols}
%\section
\appendix
%\appendix
%\title{Supplimentary material}
%\onecolumn
\renewcommand{\thesection}{\arabic{section}}
\renewcommand{\thesubsection}{\thesection.\arabic{subsection}}
\renewcommand{\thesubsubsection}{\thesubsection.\arabic{subsubsection}}
\section{Langevin equation for a self-propelled rod in the presence of temperature gradient}\label{Langevin_equation}
The over-damped Langevin equation, satisfied by a thin rod-like particle with its center of mass at $\br$ and orientation along $\bu$ is,
\begin{align}
    \begin{aligned}
        \frac{d r_i}{dt} &=v_0 u_i-D^T_{ij}\partial_j T-\Gamma^{-1}_{ij}\frac{\partial V_{ex}}{\partial x_j}+\eta_i
        \\
        \frac{d \bu}{dt} &= \bu \times \left(\bm{\zeta}+ \bm{\omega}^T -\frac{1}{\gamma_r} \pmb{\mathscr{R}} V_{ex} \right) 
        \label{langevin}
    \end{aligned}
\end{align}
$\bm{\eta}$ and $\bm{\zeta}$ are white noises with zero mean and correlations $\left\langle\zeta_i(t)\zeta_j(t')\right\rangle = \frac{2 k_b T}{\gamma_r} \delta_{ij}\delta(t-t')=2 D_r \delta_{ij}\delta(t-t')$ and $\left\langle\eta_i(t)\eta_j(t')\right\rangle = 2 k_b T \Gamma^{-1}_{ij}\delta(t-t')=2 D_{ij}\delta(t-t')$. The term $\Gamma_{ij}$ in equation \ref{langevin} is $ij$-th component of viscosity tensor, that has the form $\Gamma_{ij}=\Gamma_{\parallel}u_i u_j+\Gamma_{\perp}(\delta_{ij}-u_iu_j)$ where $\Gamma_{\parallel}$ is the viscosity felt by the particle along its orientation and $\Gamma_{\perp}$ is the viscosity felt by the particle perpendicular to the orientation of the particle. So the translational diffusion coefficient tensor has the form $D_{ij}=D_{\parallel} u_i u_j+D_{\perp}(\delta_{ij}-u_i u_j)$ with $D_{\parallel}=k_b T/\Gamma_{\parallel}$ and $D_{\perp}=k_b T/\Gamma_{\perp}$ and the rotational diffusion coefficient is $k_bT/\gamma_r$. The exact expressions of all the diffusion coefficients can be derived using hydrodynamics \cite{doi1988theory}, $D_{\perp}=D_{\parallel}/2$ and $D_r= 6 D_{\parallel}/l^2$. For our convenience, we have defined $D_{\parallel}=D$.
\par 
Each particle interacts with its neighboring particles via an excluded volume interaction $V_{ex}$,. %which is termed excluded volume interaction and its detailed expression is given in the paper. The second term present in each of the coupled equations,\ref{langevin} comes from the interaction of the particles with the temperature gradient field(check sec Thermophoresis).
\par 
Applying Ito calculus to the  coupled Langevin equation \ref{langevin}, we get the Smoluchowski equation satisfied by the phase space probability density function $\psi(\br,\bu,t)$,
\begin{align}
    \begin{aligned}
        \partial_t \psi &= -v_0 \bu\cdot\bm{\nabla}\psi+\partial_i(\psi D^T_{ij}\partial_j T)+\frac{k_b T \Gamma^{-1}_{ij}}{k_b T}\partial_i\left(\psi \partial_j V_{ex}\right)+\partial_i\partial_j D_{ij}\psi
        \\ &+D_r \pmb{\mathscr{R}}^2 \psi+\frac{1}{\gamma_r}\pmb{\mathscr{R}}(\psi\pmb{\mathscr{R}}V_{ex})-\pmb{\mathscr{R}}\cdot\bm{\omega^T}
        \\&
        = -v_0 \bu\cdot\bm{\nabla}\psi+\partial_i(\psi D^T_{ij}\partial_j T)+\frac{D_{ij}}{k_b T}\partial_i\left(\psi \partial_j V_{ex}\right)+D_{ij}\partial_i\partial_j \psi+\partial_i D_{ij}\partial_j \psi
        \\
        &+\partial_j D_{ij}\partial_i \psi+\psi\partial_i \partial_j D_{ij}+D_r \pmb{\mathscr{R}}^2 \psi+\frac{D_r}{k_bT}\pmb{\mathscr{R}}\cdot(\psi\pmb{\mathscr{R}}V_{ex})-\pmb{\mathscr{R}}\cdot\bm{\omega^T} 
    \end{aligned}
\end{align}
In general, for non-uniform temperature, $D_{ij}(\br)$ can be expanded in temperature gradient as,
\begin{align}
\begin{aligned}
    D_{ij}=k_b T(\br)\Gamma^{-1}_{ij}=k_b \Gamma^{-1}_{ij} \left(T(0)+\bm{\nabla}T\cdot\br+...\right)
\end{aligned}
\end{align}
However, for small temperature gradients, which we consider in this paper, $D_{ij}$ can be considered constant, independent of spatial variation. This leads to the following Smoluchowski equation for $\psi(\br,\bu,t)$,
\begin{align}
\nonumber
   \partial_t \psi &= -v_0 \bu\cdot\bm{\nabla}\psi+\partial_i(\psi D^T_{ij}\partial_j T)+\frac{D_{ij}}{k_b T}\partial_i\left(\psi \partial_j V_{ex}\right)+D_{ij}\partial_i\partial_j \psi
   \\
   &+D_r \pmb{\mathscr{R}}^2 \psi+\frac{D_r}{k_bT}\pmb{\mathscr{R}}\cdot(\psi\pmb{\mathscr{R}}V_{ex})-\pmb{\mathscr{R}}\cdot\bm{\omega^T} 
\end{align}
\section{Moment Expansion of concentration}\label{moment_expansion}
The concentration $\psi(\textbf{r},\hat{\textbf{u}},t)$ can be expanded as sum of irreducible tensors $T^m_{i_1,i_2,...i_m}$ which are equivalent to spherical harmonics but expressed in cartesian coordinates. The components of $T^m_{i_1,i_2,...i_m}$ are homogeneous polynomials of degree $m$ with components $u_m$ of unit vector $\hat{u}$ in dimension $d$. Each tensor is orthogonal to any other tensor.
The symmetric trace-less irreducible tensors up to 6th rank for dimension $d$ are as follows,
%\lipsum
%\begin{strip}
%\begin{widetext}
%\onecolumn
\begin{align}
\begin{aligned}
T^0 &=1
\\
T^1 _i &=u_i
\\
T^2_{ij} &=u_iu_j-\frac{\delta_{ij}}{d}
\\
T^3_{ijk} &=u_iu_ju_k-\frac{1}{d+2}[\delta_{ij}u_k+\delta_{jk}u_i+\delta_{ik}u_j]
\\ \nonumber
T^4_{ijkm} &=u_iu_ju_ku_m-\frac{1}{d+4}[\hat{Q}_{ij}\delta_{km}+\hat{Q}_{ik}\delta_{jm}+\hat{Q}_{im}\delta_{jk}
+\hat{Q}_{jk}\delta_{im}+\hat{Q}_{jm}\delta_{ik}+\hat{Q}_{k}m\delta_{ij}]
\\
&-\frac{1}{d(d+2)}\Delta_{ijkm}
\\ \nonumber
T^5_{ijkmn}& =u_iu_ju_ku_mu_n -\frac{1}{d+6}[T^3_{kmn}\delta_{ij}+...all~other ~combinations]
 \\
 &
-\frac{1}{(d+2)(d+4)}[u_i\Delta_{jkmn}+u_j\Delta_{ikmn}+u_k\Delta_{ijmn}+u_m\Delta_{ijkn}+u_n\Delta_{ijkm}]
\\
T^6_{ijkmno}&=u_iu_ju_ku_mu_nu_o+c_1[T^4_{ijkm}\delta_{no}+...all~other ~combinations]
\\
&-\frac{1}{(d+4)(d+6)}[\hat{Q}_{ij}\Delta_{kmno}+\hat{Q}_{ik}\Delta_{jmno}+\hat{Q}_{im}\Delta_{jkno}+\hat{Q}_{in}\Delta_{jkmo}\\&+\hat{Q}_{io}\Delta_{jkmn}+
\hat{Q}_{jk}\Delta_{imno}+\hat{Q}_{jm}\Delta_{ikno}+\hat{Q}_{jn}\Delta_{ikmo}+\hat{Q}_{jo}\Delta_{ikmn}+\hat{Q}_{km}\Delta_{ijno}
\\&+\hat{Q}_{kn}\Delta_{ijmo}+\hat{Q}_{ko}\Delta_{ijmn}
 +\hat{Q}_{mn}\Delta_{ijko}+\hat{Q}_{mo}\Delta_{ijkm}+\hat{Q}_{no}\Delta_{ijkm}]
\\ & -\frac{1}{d(d+2)(d+4)}[\delta_{ij}\Delta_{kmno}+\delta_{ik}\Delta_{jmno}+\delta_{im}\Delta_{jkno}
+\delta_{in}\Delta_{jkmo}+\delta_{io}\Delta_{jkmn}]
\end{aligned}
\end{align}
%\end{widetext}
%\end{strip}
%\lipsum
Where $\Delta{ijkm}=\delta_{ij}\delta_{km}+\delta_{ik}\delta_{mn}+\delta_{im}\delta_{jk}$.
%\par
The concentration $\psi(\textbf{r},\hat{\textbf{u}},t)$ can be expanded in the basis of $T^m$s as
\begin{align}\label{series}
\psi(\textbf{r},\hat{\textbf{u}},t)=\sum_{m=0}^\infty a^m_{i_1,i_2,..i_m}(\br,t)T^m_{i_1,i_2,...i_m}(\hat{u})
\end{align}
$a^m_{i_1,i_2,..,,i_m}$'s are called $m-$th moment of the concentration.
Trancating the series, (\ref{series}) to $m=2$,
\begin{align}\label{c}
\nonumber
\psi(\textbf{r},\hat{\textbf{u}},t)&=a^0(\br,t)+a^1_i(\br,t)T^1_i+a^2_{ij}(\br,t)T^2_{ij}
\end{align}

To determine $a^0,a^1_i$ and $a^2_{ij}$ we have to multiply $\psi(\textbf{r},\hat{\textbf{u}},t)$ with $T^0,T^1$ and $T^2$ tensors respectively and integrate with respect to $\hat{u
}$. Using the above expressions for the irreducible tensors and considering that they are orthogonal to each other, one can obtain the following forms of moment tensors:

 \begin{align}
 \nonumber
 a^0 &=\frac{1}{\Omega_d}\int_{\hat{u}}\psi(\textbf{r},\hat{\textbf{u}},t)=\frac{1}{\Omega_d}\rho(\br,t)
 \\
 \nonumber
 a^1_i &=\frac{d}{\Omega_d}\int_{\hat{u}}u_i \psi(\textbf{r},\hat{\textbf{u}},t)=\frac{d}{\Omega_d}P_i(\br,t)=\frac{d}{\Omega_d}\rho(\br,t)p_i(\br,t)
 \\
 \nonumber
 a^2_{ij} &= \frac{d(d+2)}{2\Omega_d}\int_{\hat{u}}\psi(\textbf{r},\hat{\textbf{u}},t)\hat{Q}_{ij}=\frac{d(d+2)}{2\Omega_d}Q_{ij}(\br,t)=\frac{d(d+2)}{2\Omega_d}\rho(\br,t)S_{ij}(\br,t)
  \end{align}
  Where $\Omega_d$ is solid angle in $d-$dimensional space and $\hat{Q}_{ij}=u_iu_j-\frac{\delta_{ij}}{d}$
  
  \par
  So, in $d-$dimension, $\psi(\textbf{r},\hat{\textbf{u}},t)$ can be written as,
  
  \begin{align}
  \nonumber
  \psi(\textbf{r},\hat{\textbf{u}},t)&=\frac{1}{\Omega_d}\Big[\rho(\br,t)+d P_i(\br,t)u_i+\frac{d(d+2)}{2} Q_{ij}(\br,t)u_iu_j-\frac{\delta_{ij}}{d}
\Big]
  \\ &=\frac{1}{\Omega_d}\rho(\br,t)\Big[1+d p_i(\br,t)u_i+\frac{d(d+2)}{2} S_{ij}(\br,t)(u_iu_j-\frac{\delta_{ij}}{d})\Big]
  \end{align}
 \section{Coarse-grained currents} 
\subsection{Translational contribution}\label{coarse-grained_detailed}
\subsubsection{Due to translational diffusion}
\begin{align}
\nonumber
D_{ij} &=Du_iu_j+D/2(\delta_{ij}-u_iu_j) 
=D/2(u_iu_j+\delta_{ij})
\\ \nonumber
J_{i} &=-\int_{\textbf{u}} D_{ij}\nabla_{j} \psi(\textbf{r},\textbf{u},t)=-\int_{\textbf{u}} \Big(\frac{D}{2}u_{i}u_{j}+\frac{D}{2}\delta_{ij} \Big) \nabla_{j}\psi
\\&
=-\frac{D}{2}\int_{\textbf{u}} \Big[(u_{i}u_{j}-\frac{\delta_{ij}}{d})+\delta_{ij}\frac{d+1}{d}\Big]\nabla_j \psi
\end{align}
Identifying the first term of the RHS as the second rand rank irreducible tensor $\hat{Q}_{ij}$,
\begin{equation}
J_{i}=-\frac{D}{2}\Big[\nabla_{j}Q_{ij}+\frac{d+1}{d}\delta_{ij}\nabla_{j}\rho \Big]
\end{equation}
\begin{align}
\nonumber
J_{ij} &=-\int_{\textbf{u}} u_{i}D_{jk}\nabla_{k} c(\textbf{r},\textbf{u},t)
=-\frac{D}{2}\Big[\int_{\textbf{u}} u_{i}u_{j}u_{k}\nabla_{k}c+\int_{\textbf{u}} u_{i}\nabla_{j}c\Big]
\end{align}
Neglecting higher-rank tensors 
\begin{align}
\nonumber
J_{ij}=-\frac{D}{2}\frac{1}{d+2}\Big[(d+3)\nabla_{j}P_i+\delta{ij}\vec{\nabla}.\vec{P}+\nabla_i P_j\Big]
\end{align}
\begin{align}
\nonumber
 J_{ijk}&= \int_{\textbf{u}}\hat{Q_{ij}}D_{km}\nabla_{m}c(\textbf{r},\textbf{u},t)
\\ \nonumber
 &=-\frac{D}{2}\int_{\textbf{u}}u_i u_j u_k u_m \nabla_m c-\frac{D}{2}\int_{\textbf{u}} u_i u_j \nabla_k c+\frac{D}{2}\frac{\delta_{ij}}{d}\int_{\textbf{u}} u_k u_m \nabla_m c +\frac{D}{2}\frac{\delta_{ij}}{d}\nabla_k \rho
\\ \nonumber
 &= -\frac{D}{2}\Big[\frac{1}{d+4}\Big((d+5)\nabla_k Q_{ij}+\nabla_j Q_{ik} + \nabla_i Q_{jk} -\frac{4}{d}\delta_{ij}\nabla_m Q_{km} + 
 \delta_{jk}\nabla_m Q_{im} +\delta_{ik}\nabla_m Q_{jm}\Big)
 \\
 &+\frac{1}{d(d+2)}\left(-\frac{2}{d}\delta_{ij}\nabla_k \rho+\delta_{jk}\nabla_i \rho+\delta_{ik}\nabla_j \rho\right)\Big]
 \end{align}
\subsubsection{Contribution due to Excluded volume interaction}
Excluded volume interaction potential
\begin{align}
\begin{aligned}
V_{ex}(\br_1,t)&=2 \nu k_b T \int dr_2\int du_2 \psi(\br_2,\bu_2,t)|\bu_1 \times \bu_2|
\int _{s_1,s_2} \delta(\br_1+\bu_1s_1-\br_2-\bu_2s_2)
\\
&=2\nu k_b T\int du_2\int _{s_1,s_2}\psi(\br_1+\bm{\xi} ,\bu_2,t)|\bu_1 \times \bu_2 |
\label{ex_vol}
\end{aligned}
\end{align}
where $\xi= |s_1\bu_1-s_2\bu_2|=|\br_2-\br_1| $.
Taylor expansion of $\psi(r_1+\xi ,\bu_2,t)$ about $\br_1$ gives,
\begin{align}
 \begin{aligned}
     \psi(\br_1+\bm{\xi} ,\bu_2,t)&=\psi(\br_1 ,\bu_2,t)+\xi_i\partial_i \psi(\br_1,\bu_2,t)
     +\frac{1}{2}\xi_i\xi_j\partial_i \partial_j \psi(\br_1,\bu_2,t)
     \label{psi_taylor}
 \end{aligned}   
\end{align}
We have considered the angle between $\bu_1$ and $\hat{u_2}$ is small. Therefore,
\begin{align}
\begin{aligned}
|\bu_1 \times \bu_2 |&=\sqrt{1-(\bu_1.\bu_2)^2}
=1-\frac{1}{2}(\bu_1.\bu_2)^2
=1-\frac{1}{2}u_{1\alpha}u_{2\alpha}u_{1\beta}u_{2\beta}
\end{aligned}
\end{align}
Putting this and Eq. \ref{psi_taylor} in Eq. \ref{ex_vol} we can get
\begin{equation}
V_{ex}(\br_1,t)= 2 \nu k_bT  l^2\Big[\rho\left(1-\frac{1}{2d}\right)-\frac{1}{2}u_{1\alpha}u_{2\beta}Q_{\alpha\beta}\Big]
\label{excludedvol}
\end{equation}
Excluded volume current density
\begin{equation}
J^{ex}_i=-\frac{D_{ij}}{k_b T}\psi \nabla_j V_{ex}
\end{equation}
coarse-grained currents
\begin{align}
\begin{aligned}
J^{ex}_i&=-\frac{D}{2}2 \nu l^2\frac{2d-1}{2d}\nabla_j \Big[\rho\big(Q_{ij}
+\frac{d+1}{2d}\delta_{ij}\rho\big)\Big]+\frac{D}{8}
\\&+2 \nu l^2 \frac{d+5}{d+4}\nabla_i (Q_{\alpha\beta}Q_{\alpha\beta})\frac{D}{2}2 \nu l^2\frac{1}{d+4}\nabla_j Q_{i\alpha}Q_{j\alpha}+\frac{D}{2}2 \nu l^2\frac{2d+3}{2(d+2)}\rho \nabla_j Q_{ij}
\\ \nonumber
J^{ex}_{ij} &=-\frac{D}{2}2 \nu l^2(1-\frac{1}{2d})\frac{1}{d+2}\Big[\delta_{ij}\vec{\nabla}\rho.\vec{P}+(d+3)P_i\nabla_j\rho+P_j\nabla_i \rho  \Big]+\frac{D}{2}2\nu l^2\frac{1}{(d+2)(d+4)}\\
& \Big[P_i\nabla_k Q_{jk}+P_j\nabla_k Q_{ik}+P_k\nabla_k Q_{ij}+\delta_{ij}P_\alpha\nabla_k Q_{k\alpha}+(d+5)P_\alpha\nabla_j Q_{i\alpha}+P_\alpha\nabla_i Q_{j\alpha}\Big]
\\ \nonumber
J^{ex}_{ijk} &= -\frac{D}{2}2\nu l^2\Big(1-\frac{1}{2d}\Big)\frac{1}{d+4}\Big[(d+5)Q_{ij}\nabla_k \rho+Q_{ki}\nabla_j \rho+Q_{jk}\nabla_i\rho+\delta_{ki}Q_{jm}\nabla_m\rho
\\&+\delta_{jk}Q_{im}\nabla_m\rho
-\frac{4}{d}\delta_{ij}Q_{km}\nabla_m\rho\Big]-\frac{D}{4}2\nu l^2\Big(1-\frac{1}{2d}\Big)\frac{1}{d(d+2)}\Big[\delta_{ki}\nabla_j+\delta_{jk}\nabla_i-\frac{2}{d}\delta_{ij}\nabla_k\Big]\rho^2
\\ \nonumber
&+\frac{D}{2}2\nu l^2\frac{1}{2(d+4)(d+6)}\nabla_m\Big[Q_{im}Q_{jk}+Q_{jm}Q_{ki}+Q_{km}Q_{ij}-\frac{6}{d}\delta_{ij}Q_{k\alpha}Q_{\alpha m}
\\&+\delta_{jk}Q_{i\alpha}Q_{\alpha m}
+\delta_{ki}Q_{j\alpha}Q_{\alpha m}\Big]
+\frac{D}{8}2\nu l^2\frac{1}{(d+4)(d+6)}\Big[-\frac{30+4d}{d}\delta_{ij}\nabla_k+\delta_{jk}\nabla_i+\delta_{ki}\nabla_j\Big]Q^2_{\alpha\beta}
\\&+\frac{D}{2}2 \nu l^2\frac{1}{d+4}\nabla_k Q_{i\alpha}Q_{j\alpha}+\frac{D}{2}2\nu l^2\frac{1}{d(d+2)(d+4)}\rho\Big[(d+5)\nabla_k Q_{ij}+\nabla_j Q_{ki}+\nabla_i Q_{jk}\Big]
\\&+
\frac{D}{2}2\nu l^2\frac{1}{d(d+2)(d+4)}\rho\nabla_m\Big[\delta_{jk}Q_{im}+\delta_{jk}Q_{im}-\frac{4}{d}\delta_{ij}Q_{km}\Big]
\end{aligned}
\end{align}
Here, we have neglected $\bm{\nabla}T$ terms as $\lvert\bm{\nabla}T\rvert$ is small.
\subsubsection{Contribution of temperature gradient}
current density due to thermophoresis 
\begin{align}
\begin{aligned}
{J_{ci}^T}(\textbf{r},\hat{u},t)&=-D^T_{ij} \nabla_j Tc(\textbf{r},\hat{u},t)
\end{aligned}
\end{align}
where $D^T_{ij}=D^{T}_{||}u_iu_j+D^T_{\perp}(\delta_{ij}-u_iu_j)$
coarse-grained current,                                                  
\begin{align}
\begin{aligned}
J^{T}_i &=-\int_{\hat{u}}\big[(D^{T}_{||}-D^T_{\perp})u_iu_j+D^T_{\perp}\delta_{ij}\big](\nabla_j T) c(\textbf{r},\hat{u},t)
\\
&=-(D^{T}_{||}-D^T_{\perp})(\nabla_j T)Q_{ij}-\frac{1}{d}\rho\nabla_i T\big[D^{T}_{||}+(d-1)D^T_\perp \big]
\\
J_{ij}^T &=-\frac{D^{T}_{||}-D^T_{\perp}}{d+2}\delta_{ij}\vec{P}.\vec{\nabla}T-\frac{D^{T}_{||}-D^T_{\perp}}{d+2}P_j\nabla_i T- \frac{D^{T}_{||}+(d+1)D^T_{\perp}}{d+2}\nabla_j T P_i
\\ 
J_{ijk}^T &=-\frac{D^{T}_{||}-D^T_{\perp}}{d+4}\Big[Q_{ik}\nabla_j T+ Q_{jk}\nabla_i T+\delta_{jk} Q_{im}\nabla_m T +\delta_{ik} Q_{jm}\nabla_m T-\frac{4}{d}\delta_{ij}Q_{km}\nabla_m T\Big] 
\\&-\frac{D^{T}_{||}+(d+3)D^T_{\perp}}{d+4}Q_{ij}\nabla_k T- \frac{D^{T}_{||}-D^T_{\perp}}{d(d+2)}\Big[\delta_{ik}\rho \nabla_j T+\delta_{jk}\rho \nabla_i T -\frac{2}{d}\delta_{ij}\rho \nabla_k T \Big]
\end{aligned}
\end{align}
\subsubsection{Due to self-propulsion}
We set the following terms for the current density,
\begin{align}
J^a_{i} &= v_0 P_{i}
\\
J^a_{ij} &= v_0\left(Q_{ij}+\frac{\delta_{ij}}{d} \rho \right)
\\
J^a_{ijk} &=\frac{v_0}{d+2}\left(-\frac{2}{d}\delta_{ij}P_k+\delta_{jk}P_i+\delta_{ki}P_j\right)
\end{align}
\subsection{Rotational contribution}
\subsubsection{Due to rotational diffusion}
\begin{align}
\nonumber
R_i^D= & \int_{\bu} u_i(\{\pmb{\mathscr{R}}.\pmb{\mathscr{J}}_D^c)
=-D_r\int_{\bu} u_i R_j^2 c 
=D_r\int_{\bu} c R_j^2 u_i=(d-1)D_r\int_{\bu} u_ic   = (d-1)D_r P_i
\\
R^D_{ij}=&\int_{\bu}\hat{Q}_{ij}(\pmb{\mathscr{R}}.\pmb{\mathscr{J}}_D^c))=-D_r\int_{\bu } \hat{Q}_{ij} R_j^2 c=2dD_r\int_{\bu}\hat{Q}_{ij} c=2d D_r Q_{ij}
\end{align}
\subsubsection{due to excluded volume }Using the simplified expression of $V_{ex}$ in equation \ref{excludedvol},
\begin{align}
\begin{aligned}
R^{ex}_i &= -D_r 2\nu l^2\frac{d}{d+2}Q_{ij}P_j
\\
R^{ex}_{ij} &= -4D_r l^2 \Big[\frac{d}{d+4}Q_{ik}Q_{jk}+\frac{1}{d+2}\rho Q_{ij}-\frac{1}{d+4}\delta_{ij}Q^2_{\alpha\beta}\Big]
\end{aligned}
\end{align}
\subsubsection{Due to temperature gradient}
\begin{align}
\begin{aligned}
R^T_i =& -D^T_{\perp}\frac{1}{d+2}\left( P_i\nabla^2 T-d P_j\partial_i\partial_j T \right)
\\
R^T_{ij} =& -D^T_{\perp}\left[\frac{2}{d+4}\left(\delta_{ij}Q_{pq}\partial_p\partial_q T+Q_{ij}\nabla^2 T\right) \right]
+D^T_{\perp}\left[\frac{d}{d+4}(Q_{ip}\partial_j\partial_p T+Q_{jp}\partial_i\partial_p T)\right]
\\&
-\frac{2D^T_{\perp}}{d(d+2)}\left[(\delta_{ij}\rho\nabla^2 T-d \rho \partial_i\partial_j T)\right]
\end{aligned}
\end{align}
\section{Dynamic equations for a constant temperature gradient}\label{field_equations}
Considering $\boldsymbol{\nabla} T= \hat{x}\partial_x T$, we get the following equations for the density, polarization vector and alignment tensor in $2-$dimensions.
\begin{align}
\begin{aligned}
    \partial_t \rho &= \frac{D}{2}\left( (\partial_x^2 -\partial_y^2)Q_{xx}+2\partial_x\partial_y Q_{xy} \right)+\frac{3D}{4}\nabla^2 \rho +D^T(1-\alpha)\left(\partial_x(Q_{xx}\partial_x T)+\partial_x Q_{xy}\partial_x T\right)
    \\ & +\frac{D^T(1+\alpha)}{2}\partial_x(\rho\partial_x T)-v_0(\partial_x P_x+\partial_y P_y)
\end{aligned}
\end{align}
\begin{align}
\begin{aligned}
    \partial_t P_x & = \frac{D}{8}\left(5 \nabla ^2 P_x+2 \partial_x (\bm{\nabla}.\textbf{P})\right)+\frac{D^T}{4}\Big((3+\alpha)\partial_x(P_x\partial_x T)+(1-\alpha)\partial_y P_y \partial_x T\Big)
    \\& -v_0\left(\partial_x Q_{xx}+\partial_y Q_{xy}+\frac{1}{2}\partial_x\rho\right)-D_r\left(P_x -\nu l^2\left(Q_{xx}P_x+Q_{xy}P_y\right)\right)
    \\ 
     \partial_t P_y & = \frac{D}{8}\left(5 \nabla ^2 P_y+2 \partial_y (\bm{\nabla}.\textbf{P})\right)+\frac{D^T}{4}\Big((1-\alpha)\partial_y(P_x\partial_x T)+(1+3\alpha)\partial_x P_y \partial_x T\Big)
    \\&-v_0\left(\partial_x Q_{xy}-\partial_y Q_{xx}+\frac{1}{2}\partial_y\rho\right) -D_r\left(P_y -\nu l^2\left(Q_{xy}P_x-Q_{xx}P_y\right)\right)
    \end{aligned}
\end{align}
\begin{align}
    \begin{aligned}
    \partial_t Q_{xx}&= \frac{3D}{4}\nabla^2 Q_{xx}+\frac{D}{16}\left(\partial_x^2-\partial_y^2\right)\rho- 4 D_r(1-\rho\frac{\nu l^2}{4})Q_{xx}+\frac{D^T(1+\alpha)}{2} \partial_x(Q_{xx}\partial_x T)
    \\&+\frac{D^T(1-\alpha)}{8}\partial_x \rho \partial_x T+D^T \frac{1-3 \alpha}{8}\rho \partial_x^2 T-\frac{v_0}{4}(\partial_x P_x- \partial_y P_y)
    \\
    \partial_t Q_{xy}&= \frac{3D}{4}\nabla^2 Q_{xy}+\frac{D}{8}\partial_x\partial_y \rho- 4 D_r(1-\rho\frac{\nu l^2}{4})Q_{xy}+\frac{D^T(1+\alpha)}{2} \partial_x(Q_{xy}\partial_xT) 
    \\&+\frac{D^T(1-\alpha)}{8}\partial_y \rho \partial_x T-\frac{v_0}{4}(\partial_y P_x+ \partial_x P_y)
    \end{aligned}
\end{align}
For a nematic state, with $S_{ij}=S(n_in_j-\delta_{ij}/2)$ and $\textbf{n}=\cos\phi \hat{x}+\sin\phi \hat{y}$. Here we have taken the dimensionless form.
 \begin{align}
  \begin{aligned}
      \frac{\cos(2\phi)}{2}\partial_t(\rho S)-\rho S \sin(2\phi)\partial_t \phi &=\frac{3}{8}\cos(2\phi) \partial_x^2 \rho S-\frac{3}{4} \sin(2\phi)(2 \partial_x \rho S\partial_x\phi+ \rho S\partial_x^2\phi)
      \\&+S^T\frac{1+\alpha}{2}\left(\frac{1}{2}\cos(2\phi)\partial_x(\rho S\partial_x T)-\rho S \sin(2\phi) \partial_x\phi\partial_x T \right)
      \\&
      +\frac{1}{16}\left(\partial_x^2 \rho+2S^T\left((1-\alpha)\partial_x\rho\partial_x T+(1-3\alpha)\rho\partial_x^2T\right)\right)
      \\&-2\left(1-\rho\frac{\nu l^2}{4}\right)\rho S \cos(2\phi)-\frac{\textrm{Pe}}{4}\partial_x P_x
      \\
      \frac{sin(2\phi)}{2}\partial_t(\rho S)+\rho S cos(2\phi)\partial_t \phi &=\frac{3}{8}\sin(2\phi) \partial_x^2 \rho S+\frac{3}{4} \cos(2\phi)(2\partial_x \rho S\partial_x\phi+ \rho S\partial_x^2\phi)
      \\& +S^T\frac{1+\alpha}{2}\left(\frac{1}{2}\sin(2\phi)\partial_x(\rho S\partial_x T)+\rho S \cos(2\phi) \partial_x\phi\partial_x T \right)
      \\&-2\left(1-\rho\frac{\nu l^2}{4}\right)\rho S \sin(2\phi)-\frac{\textrm{Pe}}{4}\partial_x P_y
  \end{aligned}
  \end{align}
  The above equations can be decoupled in terms of $S$ and $\phi$,
  \begin{align}
  \begin{aligned}
  \partial_t(\rho S) &=\frac{3}{4} \partial_x^2 \rho S+S^T\frac{1+\alpha}{2}\partial_x(\rho S\partial_x T)-4\left(1-\rho\frac{\nu l^2}{4}\right)\rho S
  \\&+\frac{\cos2 \phi}{8}\left[\partial_x^2 \rho+2S^T\left((1-\alpha)\partial_x\rho\partial_x T+(1-3\alpha)\rho\partial_x^2T\right)\right]-\frac{\textrm{Pe}}{2}\left(\cos 2\phi\partial_x P_x+\sin2\phi \partial_x P_y\right)
  \\
  \partial_t \phi &=\frac{3}{4} (2 \partial_x \rho S\partial_x\phi+ \rho S\partial_x^2\phi)-\frac{\sin 2\phi}{16}\left[\partial_x^2 \rho+2S^T\left((1-\alpha)\partial_x\rho\partial_x T+(1-3\alpha)\rho\partial_x^2T\right)\right]
  \\&+S^T\frac{1+\alpha}{2}\left(\rho S  \partial_x\phi\partial_x T \right)-\frac{\textrm{Pe}}{2}\left(\cos 2\phi\partial_x P_y-\sin2\phi \partial_x P_x\right)
  \end{aligned}
  \end{align}
\section{Fields in terms of density and temperature gradient} \label{detailed}
$S$ evolves with time as,
\begin{align}
  \begin{aligned}
      \partial_t S& = \frac{\cos 2\phi}{8}(1-2 S^2)\left(\frac{\partial_x^2\rho}{\rho}(1+2 \textrm{Pe}^2)+2 S^T(1-\alpha)c\frac{\partial_x \rho}{\rho}\right)-4\left(1-\frac{\rho}{\rho_{in}}\right) S 
      \\&+\frac{S^T c(1+\alpha)}{2}\partial_x S+\frac{3+Pe^2}{4}\left(\partial_x^2 S+2\partial_x S\frac{\partial_x\rho}{\rho}\right)-\frac{Pe^2}{4}S\frac{\partial_x^2 \rho}{\rho}  
  \end{aligned}
  \end{align}
Here we have neglected terms such as $S\partial_x^2 S$, $S\partial_x S\partial_x\rho$ and other higher-order derivative terms. For the sake of the simplicity of calculation, we neglect the excluded volume interaction term for the time being and consider that $S$ originates as the effect of thermophoresis only. 
\begin{align}
  \begin{aligned}
      \partial_t S& = \frac{\cos 2\phi}{8}(1-2 S^2)\left(\frac{\partial_x^2\rho}{\rho} (1+2Pe^2)+2 S^T(1-\alpha)\partial_x T\frac{\partial_x \rho}{\rho}\right)-4 S 
      \\&-\frac{Pe^2}{4}S\frac{\partial_x^2 \rho}{\rho}  \left(\frac{3+Pe^2}{2}\frac{\partial_x\rho}{\rho}+\frac{S^T c(1+\alpha)}{2}\right)\partial_x S
  \end{aligned}
  \end{align}
  In the hydrodynamic limit, $\partial_t S \approx 0$, and for small temperature gradient and small $S$ limit, we can neglect the term quadratic in $S$.  
  \begin{align}
      \begin{aligned}
          S&=\cos 2\phi\frac{1}{32\left(1+\frac{Pe^2}{16}\frac{\partial_x^2\rho}{\rho }\right)}\left( \frac{\partial_x^2\rho}{\rho}(1+2{\textrm{Pe}}^2)+2 S^T c(1-\alpha)\frac{\partial_x \rho}{\rho}\right)
          \\&+\left(\frac{3+\textrm{Pe}^2}{2}\frac{\partial_x\rho}{\rho}+\frac{S^T c(1+\alpha)}{2}\right)\partial_x S
      \end{aligned}
  \end{align} 
  Considering $\partial_x^2\rho<<\rho$ we have neglected that in the above expression of $S$. Next, we add $S_{ex}$ as the contribution of excluded volume interaction(sometimes called entropic effect) to the above expression of $S$ and put total $S=S_{ex}+S$ back into the equation of $J_i=0$. We neglect some terms involving higher order derivatives in fields such as $\partial_x^2\rho\partial_x\rho$, $(\partial_x\rho)^2$,
  \begin{align}
  \begin{aligned}
   & (1+2\textrm{Pe}^2) \frac{S^Tc(1-\alpha)}{16}\cos^2 2\phi\partial_x^2 \rho+ \frac{(S^Tc)^2(1-\alpha)^2}{8}\partial_x \rho
    \\ &+(3+S_{ex}\cos 2\phi+2 \textrm{Pe}^2(1+S_{ex}\cos 2\phi))\partial_x \rho +(2 S^T c (1+\alpha)+S_{ex}(1-\alpha)\cos 2\phi)\rho=0
    \end{aligned}
  \end{align}
  We neglect $\partial_x^2 \rho$ and the term with $(S^Tc)^2$ in the low temperature gradient($S^T c$) limit. 
  Hence Solving for $\rho$,
  \begin{align}
  \rho(x)=\rho(0)\exp\left(\frac{-2S^T(1+\alpha+S_{ex}\cos 2\phi(1-\alpha))c x}{3+S_{ex}\cos 2\phi+2\textrm{Pe}^2(1+S_{ex}\cos 2\phi)}\right)
  \end{align}
  As $S_{ex}<1$, we have neglected its effect on density. This approximation works very well for $\alpha\approx 1/2$ for passive rods' case($Pe=0$) and $\alpha\approx 1/2(1+\textrm{Pe}^2)$ for self-propelled active rods($Pe\neq 0$), 
  \begin{align}
      \rho(x)= \rho(0)\exp\Big[-\frac{2S^Tc(1+\alpha)x}{3+2 \textrm{Pe}^2}\Big]\approx\rho(0)\Big[1-\frac{2S^Tc(1+\alpha)x}{3+2 \textrm{Pe}^2}\Big]
  \end{align}
  So, the degree of alignment $S$ only because of temperature gradient has the form,
  \begin{align}
  S = -\cos 2\phi \frac{(S^Tc)^2(1-\alpha^2)}{8(3+2 Pe^2)}
  \end{align}
  Expression of polarization vector after using hydrodynamic approximation and taking the simplest possible form,
  \begin{align}
      p_x &= \textrm{Pe}(1+S \cos 2\phi) \frac{\partial_x \rho}{\rho}= \textrm{Pe}\left((1+S \cos 2\phi)\right)\Big(-\frac{2S^Tc(1+\alpha)}{3+2 Pe^2}\Big)
      \\
      p_y &= \textrm{Pe}(S \sin 2\phi) \frac{\partial_x \rho}{\rho}=\textrm{Pe}  S \sin 2\phi \Big(-\frac{2S^Tc(1+\alpha)}{3+2 \textrm{Pe}^2}\Big)
  \end{align}
  \subsection{Solution for steady-state $S$ with $\partial_x S$}
   Solving Eq.(45) with the boundary condition $S_{st}(0)=S_0$, 
   \begin{align*}
      S_{st}(x)&= \frac{\exp \left(-\frac{2 \left(\epsilon  \left(2 Pe^2+3\right)-2 (\alpha +1)S^T c (\epsilon +1) x\right)^2}{3 (\alpha +1)^2 (S^Tc)^2 (\epsilon +1)}\right) }{24 \sqrt{\epsilon +1}}\left[24 \sqrt{\epsilon +1}S_0 \exp \left(\frac{2 \epsilon ^2 \left(2 Pe^2+3\right)^2}{3 (\alpha +1)^2 (S^Tc)^2 (\epsilon +1)}\right)\right.
      \\&\left.+\sqrt{6 \pi } S^Tc(1-\alpha ) \text{Erfi}\left(\frac{\sqrt{\frac{2}{3}} \epsilon  \left(2 \textrm{Pe}^2+3\right)}{(\alpha +1) S^Tc \sqrt{\epsilon +1}}\right)\right.
      \\
      &\left.+\sqrt{6 \pi } S^Tc (1-\alpha)  \text{Erfi}\left(\frac{\sqrt{\frac{2}{3}} \left(2 (\alpha +1)S^T c (\epsilon +1) x-\epsilon  \left(2 Pe^2+3\right)\right)}{(\alpha +1) S^Tc \sqrt{\epsilon +1}}\right)\right]
   \end{align*}
  Taylor series expansion of the error function about $x=0$,
   \begin{align}
       \text{Erfi(x)}=\frac{2 x}{\sqrt{\pi }}+\frac{2 x^3}{3 \sqrt{\pi }}+\frac{x^5}{5 \sqrt{\pi }}+\frac{x^7}{21 \sqrt{\pi }}+...
   \end{align}
\end{document}